\documentclass[a4paper,11pt]{article}
\pdfoutput=1 

\usepackage{jinstpub} 
\let\ifpdf\relax
\usepackage{color}
\usepackage[normalem]{ulem}
\usepackage{lineno}
\usepackage{graphicx}
\usepackage{subcaption}
\usepackage{mwe}
\usepackage{amsmath}
\usepackage{gensymb}
\usepackage[symbol]{footmisc}
\usepackage{siunitx}
\usepackage{newtxmath,newtxtext}
\usepackage{float}
\usepackage{lineno}
\usepackage{overpic}
\usepackage{xcolor}

\newlength{\figwidth}
\newlength{\fighalfwidth}
\newcommand{\nue}{$\nu_{e}$}

\begin{document}

\title{\boldmath \center \LARGE 
Wire-Cell 3D Pattern Recognition Techniques for Neutrino Event Reconstruction in Large LArTPCs: \\
Algorithm Description and Quantitative Evaluation with MicroBooNE Simulation }

\date{}

\author[gg]{P.~Abratenko}
\author[n]{R.~An}
\author[d]{J.~Anthony}
\author[r]{L.~Arellano}
\author[ff]{J.~Asaadi}
\author[dd]{A.~Ashkenazi}
\author[k]{S.~Balasubramanian}
\author[k]{B.~Baller}
\author[t]{C.~Barnes}
\author[w]{G.~Barr}
% ...

% ...
\affiliation[d]{University of Cambridge, Cambridge CB3 0HE, United Kingdom}
% ...
\affiliation[k]{Fermi National Accelerator Laboratory (FNAL), Batavia, IL 60510, USA}
% ...
\affiliation[n]{Illinois Institute of Technology (IIT), Chicago, IL 60616, USA}
% ...
\affiliation[r]{The University of Manchester, Manchester M13 9PL, United Kingdom}
% ...
\affiliation[t]{University of Michigan, Ann Arbor, MI, 48109, USA}
% ...
\affiliation[w]{University of Oxford, Oxford OX1 3RH, United Kingdom}
% ...
\affiliation[dd]{Tel Aviv University, Tel Aviv, Israel, 69978}
% ...
\affiliation[ff]{University of Texas, Arlington, TX, 76019, USA}
\affiliation[gg]{Tufts University, Medford, MA, 02155, USA}
% ...

  \emailAdd{microboone\_info@fnal.gov}
\date{}





\abstract{
Wire-Cell is a 3D event reconstruction package for liquid argon time projection chambers. Through geometry, time, and drifted charge from multiple readout wire planes, 3D space points with associated charge are reconstructed prior to the pattern recognition stage.
Pattern recognition techniques, including  track trajectory and $dQ/dx$ (ionization charge per unit length) fitting, 3D neutrino vertex fitting, track and shower separation, particle-level clustering, and particle identification are then applied on these 3D space points as well as the original 2D projection measurements.
A deep neural network is developed to enhance the reconstruction of the neutrino interaction vertex.
Compared to traditional algorithms, the deep neural network boosts the vertex efficiency by a relative 30\% for charged-current $\nu_e$ interactions.
This pattern recognition achieves 80-90\% reconstruction efficiencies for primary leptons, 
after a 65.8\% (72.9\%) vertex efficiency for charged-current $\nu_e$ ($\nu_\mu$) interactions.
Based on the resulting reconstructed particles and their kinematics, we also achieve 15-20\% energy reconstruction resolutions for charged-current neutrino interactions.
}


\maketitle
  
\clearpage

\section{Introduction}~\label{sec:introduction}
The liquid argon time projection chamber (LArTPC)~\cite{rubbia77,Chen:1976pp,willis74,Nygren:1976fe} is a three-dimensional tracking calorimeter widely used in neutrino physics~\cite{Amerio:2004ze,Anderson:2011ce,Acciarri:2016smi,Cavanna:2014iqa,Berns:2013usa}. Compared to water Cerenkov and liquid scintillator technologies, LArTPCs
are highly effective in differentiating electrons and photons in neutrino interactions through identifying the gap between photon conversion vertex and neutrino interaction vertex and measuring the $dE/dx$ (ionization energy loss per unit length) in the first few centimeters of electron or photon electromagnetic cascade (EM shower). This capability is advantageous for detecting $\nu_e$ charged-current interactions with high signal efficiencies and low background contamination. Utilizing this unique capability of LArTPCs, the MicroBooNE experiment~\cite{Acciarri:2016smi} aims to understand the nature of the low-energy excess of $\nu_e$-like events observed in the MiniBooNE experiment~\cite{Aguilar-Arevalo:2012fmn} and to measure neutrino-argon interaction cross sections in various inclusive and exclusive channels.

\begin{figure}[H]
  \centering
  \includegraphics[width=0.99\textwidth]{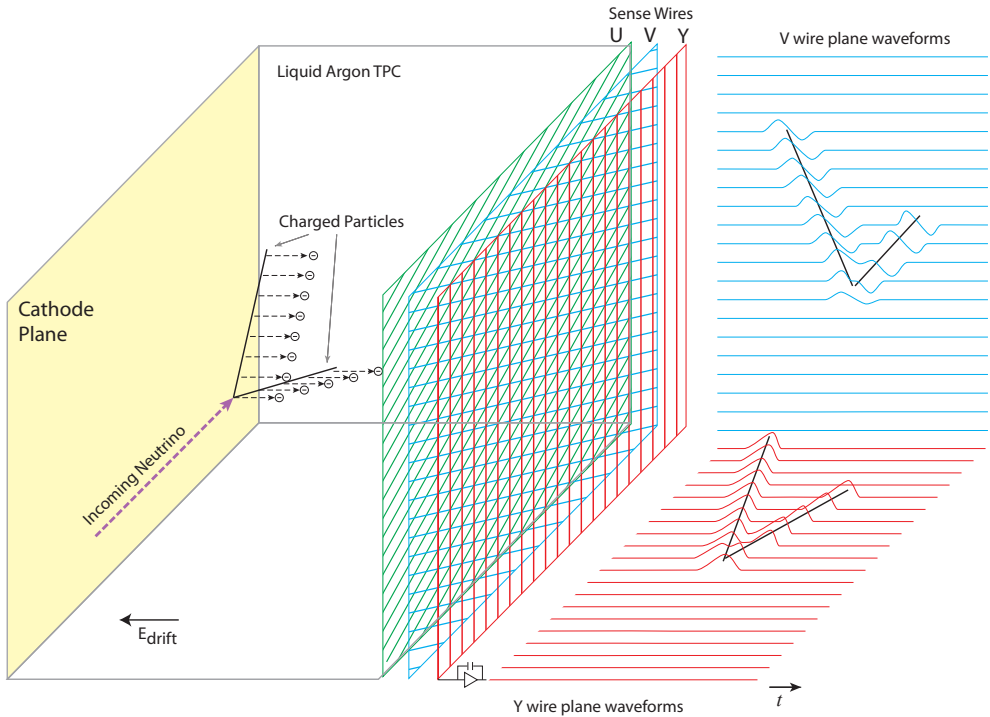}
  \caption{Illustration of signal formation in the MicroBooNE three-plane LArTPC, depicting the V plane (second induction plane) and Y plane (collection plane) TPC wire signals on the right of the image~\cite{Acciarri:2016smi}.}
  \label{fig:LArTPC_Concept}
\end{figure}

Figure ~\ref{fig:LArTPC_Concept} illustrates the MicroBooNE LArTPC, which utilizes three readout wire planes at the anode to obtain signals induced from drifting ionization electrons produced from neutrino interactions within the active volume of the detector. Drifting ionization electrons first pass by the two induction planes (the “U plane” and “V plane”) and are finally collected by the collection plane (the “Y plane”). Being one of the first large LArTPCs constructed, MicroBooNE also bears important responsibilities in the research and development of the LArTPC technology, which is key to many future experiments~\cite{Antonello:2015lea,Abi:2020wmh}. In particular, MicroBooNE pioneers the use of TPC cold electronics~\cite{Radeka:2011zz}, which significantly reduce electronic noise~\cite{Acciarri:2017sde}. Paired with enhanced TPC signal processing procedures~\cite{Adams:2018dra,Adams:2018gbi}, the cold electronics lead to a much improved performance of the induction anode wire planes. Good matching of reconstructed ionization charge between the induction and collection wire planes is demonstrated for the first time~\cite{Adams:2018dra,Adams:2018gbi}. The significantly enhanced performance of the induction wire planes provides a solid foundation in applying the Wire-Cell (a software package~\cite{Qian:2018qbv,Abratenko:2020hpp}) reconstruction to the MicroBooNE data.

\begin{figure}[H]
  \centering
  \includegraphics[width=0.65\textwidth]{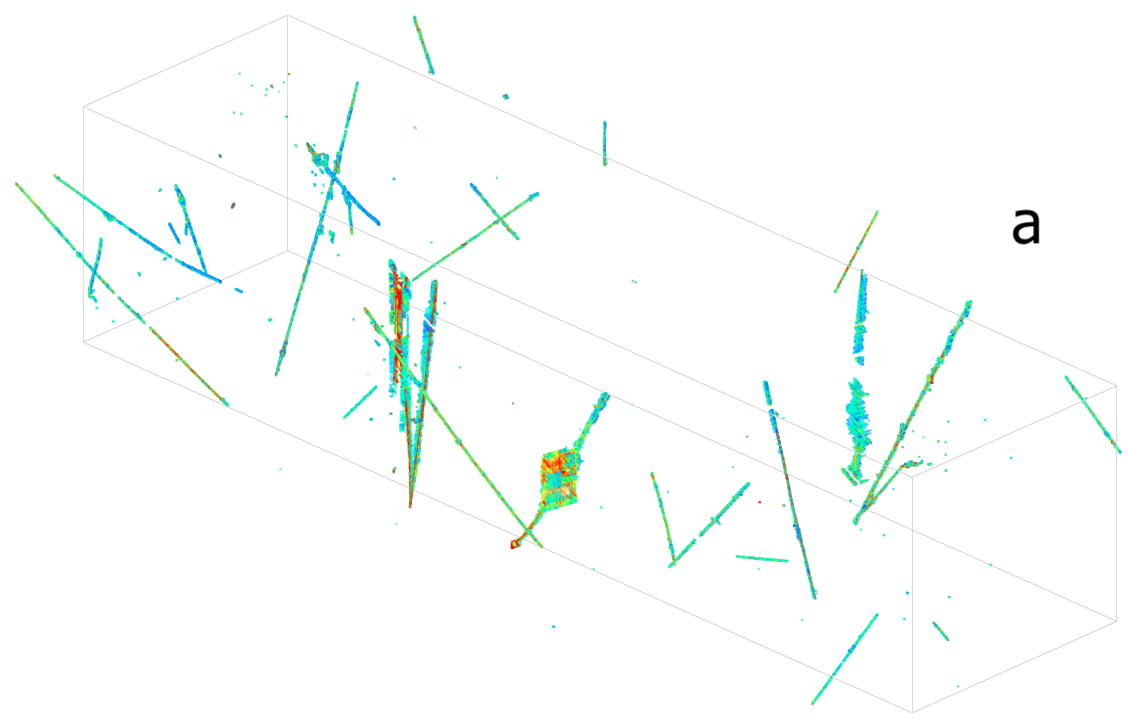}
  \includegraphics[width=0.65\textwidth]{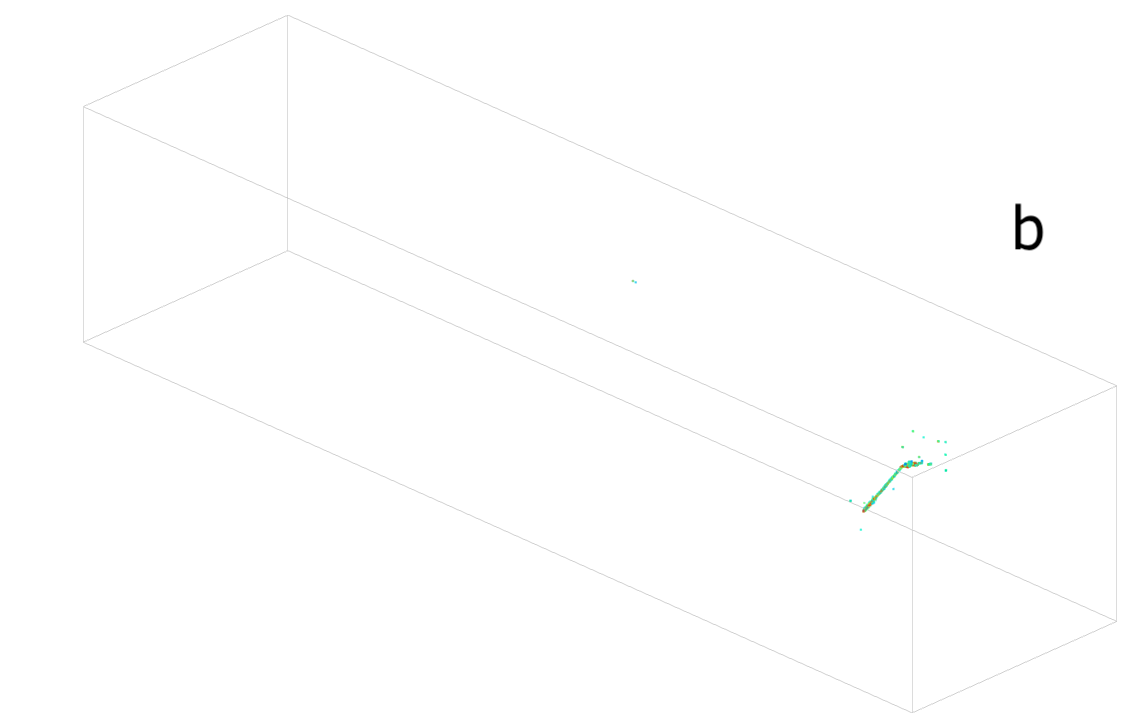}
  \put(-180, 180){MicroBooNE data}
  \put(-180, 166){(run 5384, event 2561)}
  \caption{Example event displays of the Wire-Cell 3D imaging results before (panel a) and after (panel b) generic neutrino selection. Most TPC activities in panel a are identified as cosmic rays and removed in panel b. What remains are likely associated with a neutrino interaction and will be the inputs for the Wire-Cell pattern recognition algorithms described in this paper.}
  \label{fig:generic-nu-selection}
\end{figure}

Wire-Cell imaging~\cite{Qian:2018qbv,Abratenko:2020hpp} is a 3D image reconstruction method developed by the MicroBooNE collaboration.
The main idea of Wire-Cell imaging is to reconstruct 3D images in a tomographic way without involving heuristic topological assumptions (e.g. track or shower) prior to the pattern recognition stage.
Using time and geometry information, Wire-Cell constructs the 3D space points given the observed 2D images (wire vs. time)~\cite{Qian:2018qbv}.
Afterwards, Wire-Cell utilizes the charge-matching constraint and the sparsity condition to reduce ambiguities in 3D space points introduced by the wire readout.
The charge-matching constraint requires that the same amount of reconstructed ionization charge should be seen by each (induction or collection) wire plane locally.
The distribution of ionization electrons in space is expected to be sparse, typically occupying less than 10\% of the local bounding volume that contains the activities, for any physical signals~\cite{Abratenko:2020hpp}.
This sparsity condition is applied through the compressed sensing technique~\cite{cs} via the coordinate descent algorithm~\cite{coordesc}.
The end results of Wire-Cell imaging are 3D space points with associated charges.
Figure~\ref{fig:generic-nu-selection}a shows an example of the Wire-Cell imaging result.
We note that in dealing with imperfect detector performance in MicroBooNE (e.g. dead channels)~\cite{Acciarri:2017sde}, special algorithms, including two-wire-plane tiling and deghosting, are necessary to retain the quality of 3D images.
Here, ``tiling'' is a technique that correlates 2D projections from multiple wire planes to obtain 3D information; and ``deghosting'' is a technique that removes ghost energy depositions caused by ambiguities in the 2D to 3D reconstruction~\cite{Abratenko:2020hpp}.

In the MicroBooNE LArTPC, well-reconstructed 3D images form a good foundation for the neutrino interaction selection~\cite{Abratenko:2020sxa,Abratenko:2021bzb}, which is a challenging task for LArTPCs operating on the surface because of the numerous traversing cosmic-ray muons. A set of clustering techniques on the 3D space points~\cite{Abratenko:2020hpp} was developed to separate neutrino interactions from cosmic-ray muons.
A many-to-many TPC-charge to PMT-light matching algorithm~\cite{Abratenko:2020hpp} is applied to match isolated TPC clusters to patterns of the PMT light signals, so that the neutrino interaction candidates in coincidence with the Booster Neutrino Beam (BNB) spill ($\sim$ 1.6 $\mu$s long) can be selected.
Here, a ``TPC cluster'' is defined to be a set of space points that are connected in space, given that the LArTPC is a fully active tracking calorimeter. In addition, the links between the 3D space points and their 2D projections are also kept.
To further reject cosmic muon backgrounds that occur in random coincidence with the BNB spill, a selection based on the TPC geometry is applied to reject cosmic-ray muons passing through the active TPC volume. The cosmic-ray muons that stop inside the TPC active volume can be determined by identifying the direction of the muon track. A 3D trajectory and $dQ/dx$ (ionization charge loss per unit length) fitting procedure~\cite{Abratenko:2021bzb} was developed to identify Bragg peaks and reject those stopped muons. Note that the measurable $dQ/dx$ is directly related to the previously mentioned $dE/dx$ (energy loss per unit length).
These tools are used to form a generic neutrino selection, which aims at rejecting cosmic-ray muon background while retaining high efficiencies for all kinds of neutrino interactions. 
This generic neutrino selection achieves 80\% and 88\% efficiencies for inclusive $\nu_\mu$ charged-current ($\nu_\mu$CC) and $\nu_e$ charged-current ($\nu_e$CC) interactions with a 14.9\% cosmic-ray background contamination in the selected events~\cite{Abratenko:2020sxa}.
The generic neutrino selection serves as a good basis to further apply pattern recognitions and identify various neutrino flavors with high efficiencies and purities.
Figure~\ref{fig:generic-nu-selection}b shows an example of the Wire-Cell generic neutrino selection result. In addition to the removal of cosmic-ray activity, about 80\% (90\%) of the neutrino charged-current interactions have greater than 80\% (70\%) of their energy deposits reconstructed and included in the resulting 3D TPC clusters~\cite{Abratenko:2020hpp}. This provides a solid foundation to the subsequent 3D pattern recognition. 

\begin{figure}[H]
  \centering
  \includegraphics[width=0.99\textwidth]{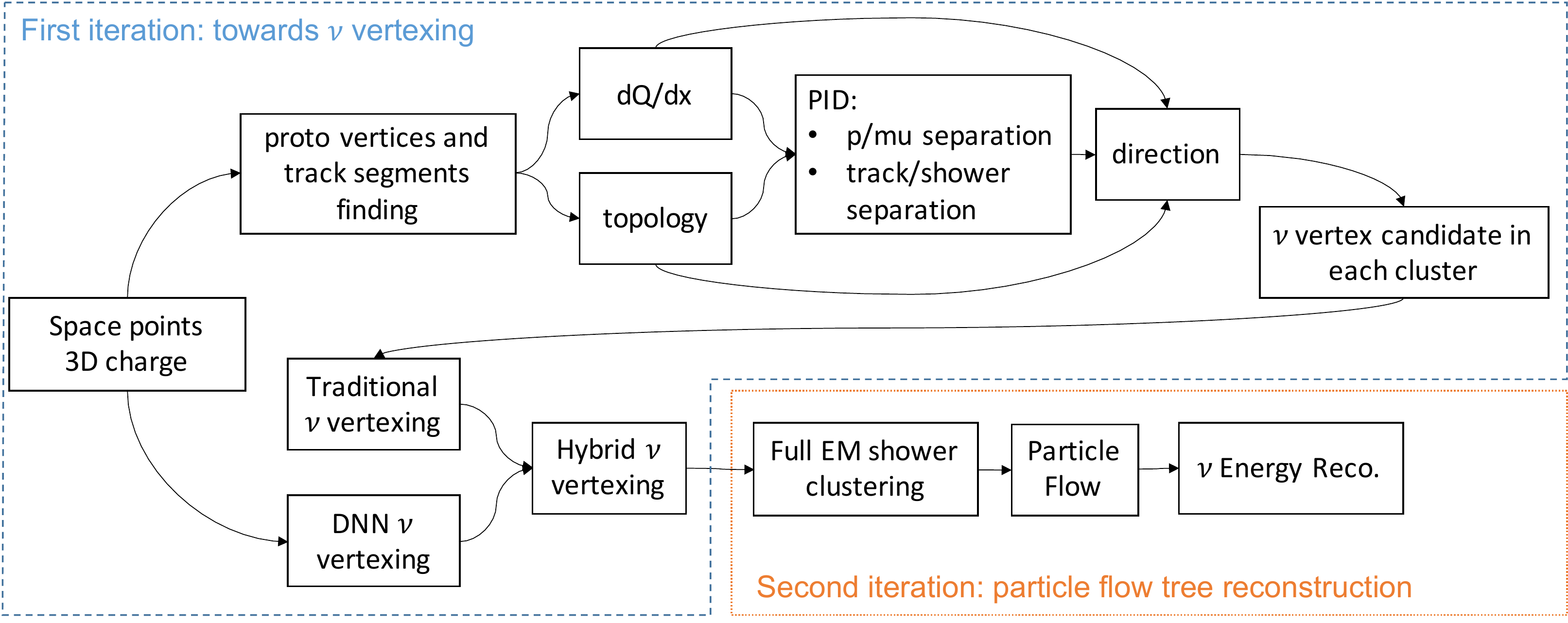}
  \caption{Overview of the Wire-Cell pattern recognition procedure. The blue dashed line indicates steps of the first iteration towards neutrino vertexing; the orange dotted line indicates steps towards particle flow tree reconstruction with a reconstructed neutrino vertex. Please note, intermediate objects reconstructed in the first iteration, e.g. track segments and initial showers, are also used in the second iteration, in addition to the reconstructed neutrino vertex. 
  More details can be found in the text.}
  \label{fig:overview_pattern_recognition}
\end{figure}

This paper summarizes the pattern recognition techniques developed and applied in Wire-Cell 
for the high-performance inclusive $\nu_e$CC and $\nu_\mu$CC event selections~\cite{WC_eLEE, WC_numuXS}.
Both traditional algorithms (non-machine-learning algorithms)
and machine-learning algorithms are used in this pattern recognition procedure.
Some of the basic tools---the track trajectory and $dQ/dx$ fitting, for example---are improved versions of the techniques developed for generic neutrino detection~\cite{Abratenko:2020sxa,Abratenko:2021bzb}. This fitting algorithm is expanded to fit multiple tracks and multiple vertices rather than fitting a single track. Figure~\ref{fig:overview_pattern_recognition} shows the overall flow of the Wire-Cell pattern recognition, which consists of two iterations.
In the first iteration, we focus on neutrino vertex identification.
An identified neutrino vertex location significantly reduces the possible solution space for other pattern recognition tasks.
Here, solution space includes the origins and directions of particles as well as the mother-daughter relationships between them.
In the second iteration, full particle flow trees are reconstructed based on the neutrino vertex locations and other intermediate objects reconstructed in the first iteration, e.g. track segments and initial showers.
Here, a particle flow tree is a tree structure that contains the mother-daughter relationship for all particles in a set.
At the beginning of the first iteration, vertices are defined by searching for kinks and splits in the reconstructed 3D images (section~\ref{sec:multi_track_fitting}). With vertices determined, track segments between vertices are defined. A 3D vertex fitting technique (section~\ref{sec:vertex_fit}) can then be used to refine the 3D vertex and connected track trajectories. Particle identification (PID) is subsequently performed on segments using $dQ/dx$ and event topology information (section~\ref{sec:track_shower_sep}). Event topology information is primarily targeted toward electromagnetic (EM) shower identification (i.e.  track/shower topology separation). Using the PID information, directions of particles can be determined in many cases.
The neutrino interaction vertex is determined using the reconstructed particle directions and later improved with a deep neutral network (DNN, section~\ref{sec:dl_nu_vtx}).
With the neutrino interaction vertex reconstructed in the first iteration, EM showers can be fully clustered (section~\ref{sec:em_clustering}), and often contain several separated sub-clusters. Finally, a particle flow tree is reconstructed including the $\pi^{0}$s reconstructed from EM showers (section~\ref{sec:pio_reco}).

\begin{figure}[!htp]
  \centering
  \includegraphics[width=0.7\textwidth]{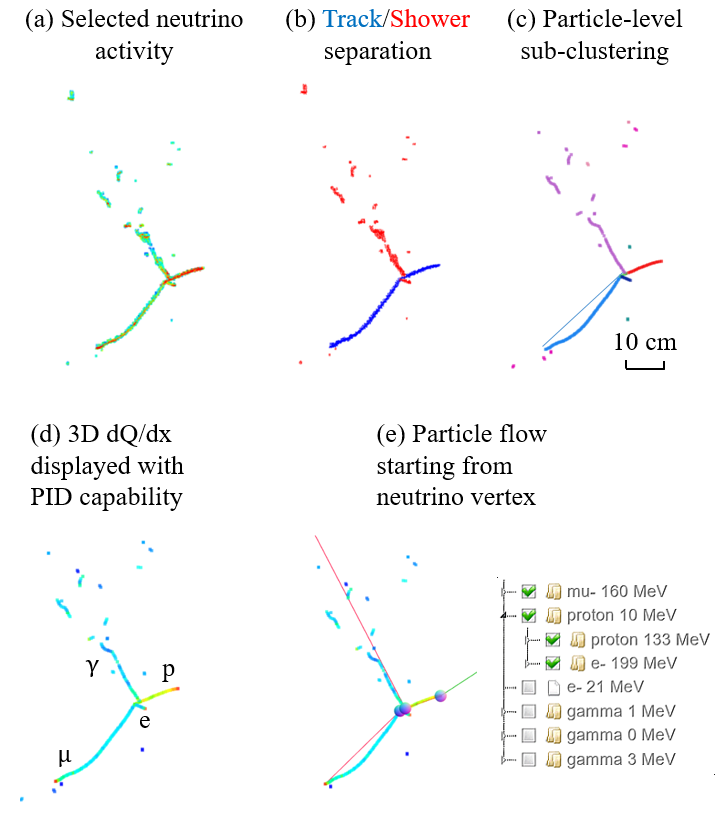}
  \put(-380, 200){MicroBooNE data}
  \put(-380, 185){(run 5384, event 2561)}
  \caption{
  Displays of the Wire-Cell pattern recognition results at different stages. (a) In-beam candidate neutrino cluster selected by the generic neutrino selection. The color scale represents the reconstructed charge associated with each space point. (b) Identified tracks and EM showers, which are displayed in blue and red, respectively. (c) Identified particles (or track segments), which are displayed in different colors. (d) Fitted $dQ/dx$ associated with each piece ($\sim$6~mm) along the trajectories. The blue, cyan, green, yellow, and red colors roughly correspond to 1/3, 1, 2, 3, and 4 times the $dQ/dx$ of a minimum ionizing particle (MIP). The short track labeled as electron is likely to be a mis-reconstructed pion here. The EM shower (listed as electron in the particle flow) is 
  labeled as a $\gamma$, given it is emitted by a proton track. It could be one of the 
  two $\gamma$s from a produced $\pi^0$. This $\gamma$ is likely converted to electron-positron pair immediately. (e) Reconstructed particle flow starting from the primary neutrino interaction vertex. The folder icon indicates the corresponding particle has reconstructed daughter particle(s), while the file icon indicates the corresponding particle has no reconstructed daughter particle(s). Particles with check marks in the check-boxes are shown in panel (e) with rainbow circles marking the starting positions and line segments marking the starting and ending positions.
  A web based event display (BEE) for this example can be found here: \url{https://www.phy.bnl.gov/twister/bee/set/uboone/reco/2021-09/lee/event/1/}.
  }
  \label{fig:illustration_pattern_recognition}
\end{figure}

Figure~\ref{fig:illustration_pattern_recognition} shows an example of the pattern recognition at different stages from a neutrino interaction event in the MicroBooNE data. 
Colors in panels (d) and (e) represent the magnitude of the reconstructed $dQ/dx$, which is especially important for PID and determining the directions of particle trajectories.
As we currently do not differentiate particles and anti-particles, in the legend, the charged particles can also refer to their antiparticles, e.g. $e^-$ can refer to both $e^-$ or $e^+$, while the final state charged particles from neutrino mode interactions are mostly negatively charged leptons and positively charged pions.

The traditional algorithms in the Wire-Cell pattern recognition introduced in this article are developed using hundreds of MicroBooNE BNB data events, while the training of the DNN neutrino vertexing and the evaluation of the performance of this pattern recognition are performed using high-statistics MicroBooNE Monte-Carlo (MC) simulation events.
E.g. in this paper, the DNN neutrino vertexing model is trained with 48000 selected $\nu_e$ charged-current events which leads to significant performance boost.
The MicroBooNE MC simulates neutrino interactions using a flux model following ref.~\cite{AguilarArevalo:2008yp} and using the GENIE v3.0.6 neutrino event generator with a MicroBooNE tune~\cite{GENIE:2021npt, uboone_genie_tune}. The particles produced in this neutrino interaction are processed through a realistic detector response model which takes the energy depositions from a Geant4 simulation~\cite{Agostinelli:2002hh}, determines the ionization charge considering recombination~\cite{Jaskolski:2011qja, Acciarri:2013met}, drifts them through the TPC electric field including the space charge effect~\cite{Adams:2019qrr, Abratenko:2020bbx}, diffuses the charge~\cite{Li:2015rqa}, and calculates the resulting signal on the wires using a sophisticated electric field response~\cite{Adams:2018dra}.
This result is in good agreement between data and MC at a fundamental level~\cite{Adams:2018gbi}.
Moreover, the MicroBooNE MC simulation used in this analysis adopts a scheme of overlaying the simulated neutrino interactions with the dedicated off-beam data that is randomly triggered when no neutrino beam spill is received, therefore the systematic uncertainties occurring in the simulation of readout noise, time-invariant detector response, and cosmic-ray backgrounds are avoided or largely reduced. Such MC simulation is referred to as overlay MC.   
This paper is organized as follows. Section~\ref{sec:pr_to_vertex} describes algorithms of the first iteration towards the identification of the neutrino vertex. Follow up steps towards full particle flow tree reconstruction are introduced in section~\ref{sec:pr_from_vertex}. Neutrino energy reconstruction is described in section~\ref{sec:energy_reco}. The conclusions are found in section~\ref{sec:summary}.

\section{Initial Iteration towards Neutrino Vertex Identification}~\label{sec:pr_to_vertex}
A key step of the neutrino event reconstruction with LArTPCs is to identify the neutrino 
interaction vertex location.
In this section, we describe two neutrino vertex identification algorithms. The first algorithm is a non-machine-learning one, utilizing particle direction information and event topology information. The second one is a machine-learning based algorithm, utilizing a sparse regressional segmentation neural network.
The final neutrino vertex position is determined by combining results from both algorithms.
As described in section~\ref{sec:introduction}, to obtain the particle direction and event topology information, dedicated algorithms for track segment finding, $dQ/dx$ fitting and track shower separation are performed.
These algorithms are introduced in this section (sub-section~\ref{sec:multi_track_fitting}, \ref{sec:vertex_fit} and \ref{sec:track_shower_sep}).
Their results are not only useful for the neutrino vertex identification but also the full particle flow reconstruction which will be introduced in section~\ref{sec:pr_from_vertex}.

\subsection{Determination of Vertices and Track Segments }~\label{sec:multi_track_fitting}
\begin{figure}[H]
  \centering
  \includegraphics[width=0.99\textwidth]{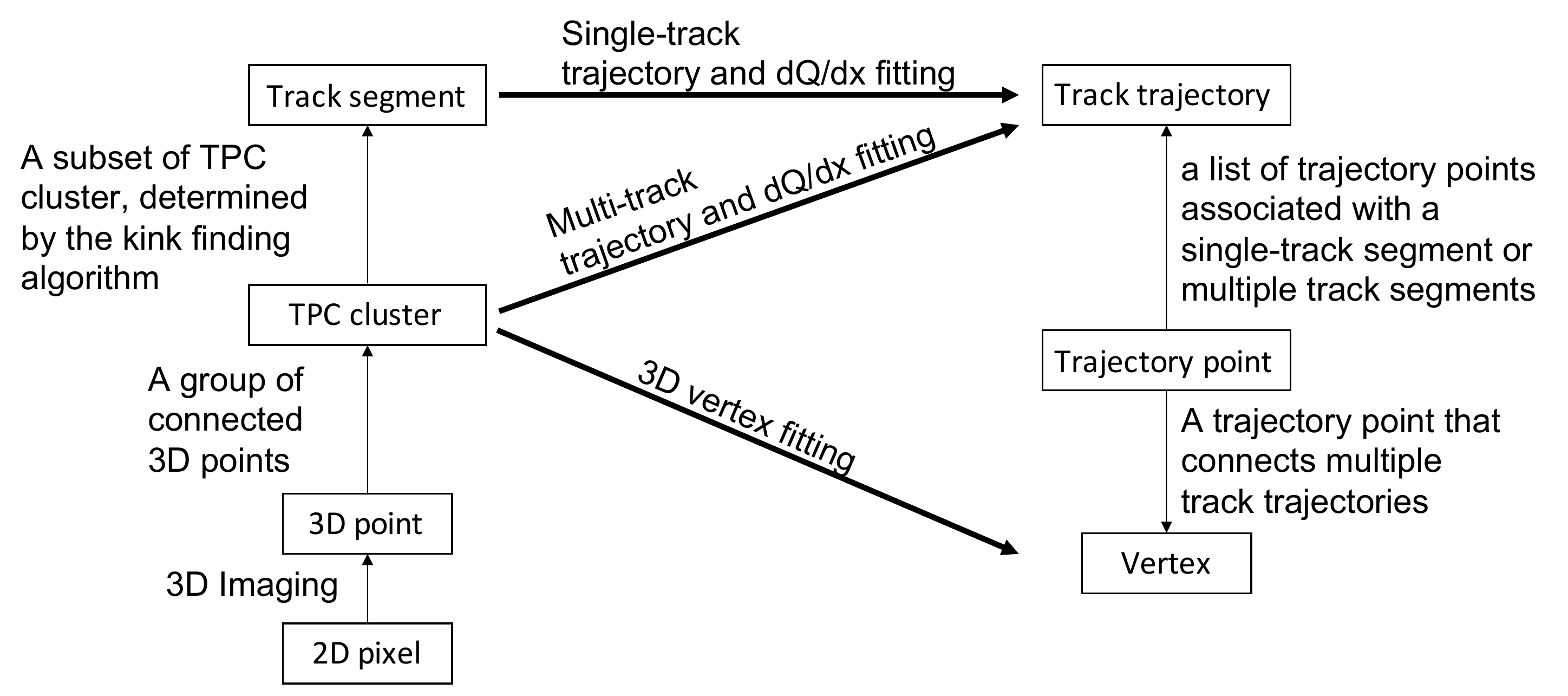}
  \caption{
    This diagram illustrates relationships between key concepts used in the multi-track trajectory and $dQ/dx$ fitting algorithm. More details about the ``3D Imaging'',  the ``Single-track trajectory and $dQ/dx$ fitting'' and the ``3D vertex fitting'' algorithms could be found in~\cite{Abratenko:2020hpp},~\cite{Abratenko:2021bzb} and section~\ref{sec:vertex_fit} respectively.
  }
  \label{fig:multi-track-fitting-flowchart}
\end{figure}

In Wire-Cell, a \textit{track segment} is defined as a subset of a TPC cluster including a set of 3D space points and associated 2D (projection) pixels with deposited charge information.
Based on a track segment, the best-fit 3D trajectory points with about 6 mm spacing are obtained by minimizing a charge-weighted distance.
That list of trajectory points is defined as a \textit{track trajectory}.
A key technique in determining vertices and track segments is the multi-track trajectory and $dQ/dx$ fitting, which is an expansion of the single-track trajectory and $dQ/dx$ fitting technique used in the generic neutrino selection~\cite{Abratenko:2021bzb}.
Figure~\ref{fig:multi-track-fitting-flowchart} illustrates the relationships between key concepts used in the multi-track trajectory and $dQ/dx$ fitting algorithm.

%

The single-track fitting technique development was inspired by the projection matching algorithm~\cite{antonello:2012hu}.
First, an initial seed of the track trajectory is obtained by constructing a Steiner-tree graph~\cite{Steiner} from the 3D points in the cluster and finding the shortest paths between extreme points.
The Steiner-tree ensures that points associated with the largest charges are included in the initial seed.
Then, the best-fit 3D trajectory points with about 6~mm spacing are obtained by minimizing a charge-weighted distance, which is constructed to compare the 2D measurements in time-versus-wire views from the three wire planes to the predictions given a 3D trajectory point.
A numerical solver for large linear systems (BiCGSTAB~\cite{BiCGSTAB}) is utilized to perform a minimization. 
With the trajectory determined, the $dQ/dx$ associated with each trajectory point can be obtained by minimizing the squared difference 
between the reconstructed and predicted ionization charge. 
A parameterized model is used to predict the measured charge taking into account the diffusion of ionization electrons during transportation and the smearing of the charge distribution in the signal processing.
This two-step procedure is adopted to avoid a nonlinear fitting process, ensuring the stability of the fit. 
Regularization on smoothness is included to further improve the $dQ/dx$ fitting performance. 
More details about this single-track fitting technique can be found in Ref.~\cite{Abratenko:2021bzb}.

\begin{figure}[H]
  \centering
  \includegraphics[width=0.99\textwidth]{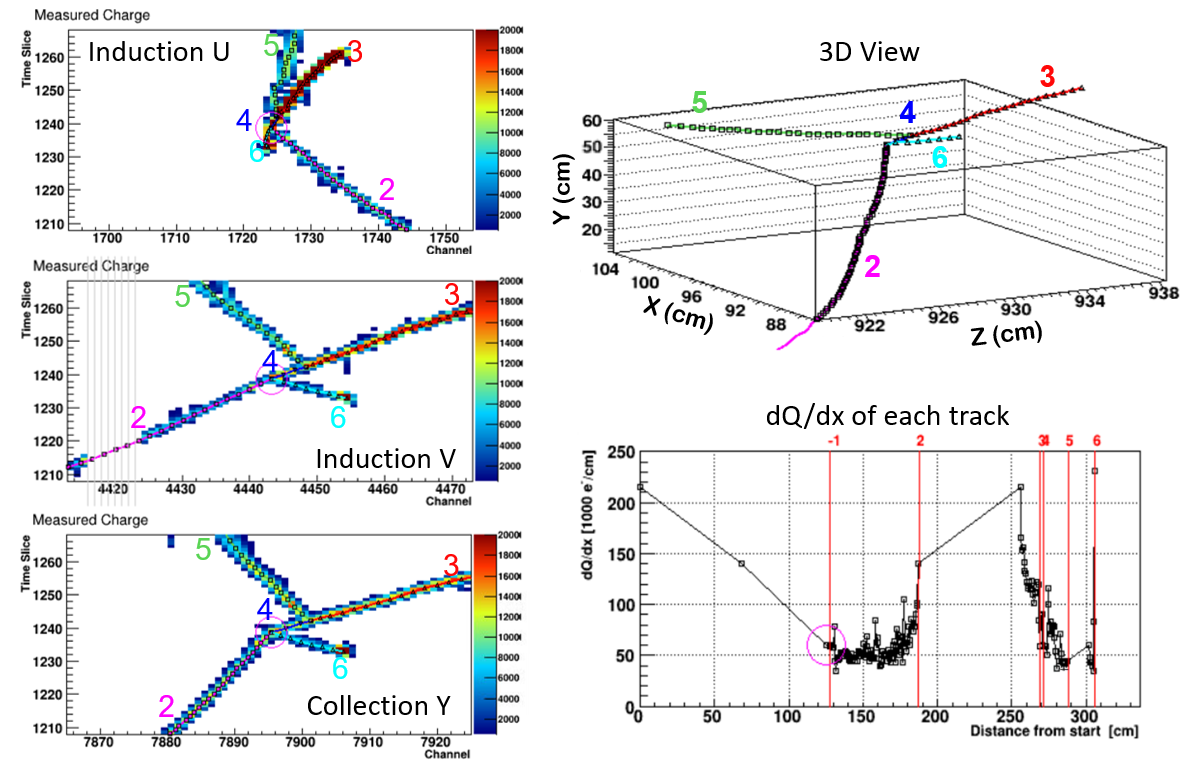}
  \put(-300, 275){MicroBooNE data (run 5384, event 2561)}
  \caption{
    Example display of the results from multi-track trajectory and $dQ/dx$ fitting.
    The three panels on the left show the reconstructed ionization charge on different planes (two induction planes and one collection plane) overlaid with the projection of the best-fit track trajectories.
    The track trajectories are color coded and labeled with numbers of the same color.
    The top right panel shows the 3D view of the best-fit track trajectories.
    The bottom right panel shows the best-fit $dQ/dx$ for all track trajectories.
    $dQ/dx$ distribution for each stopped track can be used to determine the direction of the track according to the Bragg peak.
    The color coding can be used to link the same trajectories in 2D and 3D views. Also using the color coded numbers, the trajectories can be further linked to the $dQ/dx$ fitting in the bottom right panel.
    The gray lines in the induction V view represent non-functional channels.
    In the induction U view, short track number 6 is highly compacted because of the nature of 2D projection.
  }
  \label{fig:multi-track-fitting}
\end{figure}

Moving towards the new multi-track fitting algorithm, major improvements include:
\begin{itemize}
\item The new algorithm handles simultaneous fitting of multiple track segments in 3D. 
In addition to the trajectory points that only belong to a single track segment, a new type of data product, vertex, is added. A vertex can be connected to one or more track segments.
\item The distance between two adjacent trajectory points is regulated to be as close
to 0.6~cm (twice the MicroBooNE TPC wire pitch) as possible, which makes the $dQ/dx$ fitting easier. The more uniform step size also has advantages in particle identification (PID) given the distribution of $dE/dx$ depending on the step size ($dx$).
\item Grouping of the 2D pixels in the trajectory fitting is improved to better handle multiple track segments at once. The determination of the 2D pixel group for each track segment takes into account the associated 3D space points with 2D pixels. The 3D space points are much easier to cluster with the closest track trajectory, while the 2D pixels suffer more from ambiguities because they are projections.
\item A dedicated fitting algorithm to determine the vertex position in 3D is introduced. See more detailed discussions in section~\ref{sec:vertex_fit}.
\end{itemize}
An iterative approach is used to determine track segments and vertices in a TPC cluster, 
which allows one to systematically examine the TPC cluster given its various possible
topologies.
This approach is described below:
\begin{itemize}
\item Two extreme 3D space points (a pair of end points with the largest distance between them) are found on the TPC cluster, and a single track trajectory is fitted to connect the two extreme points.
\item An algorithm is run to search for kinks\footnote{sharp turns with local angle > 25\degree} along the track trajectory from one end to the other.
When found, the track trajectory is broken at the kink vertex into separate segments, then the track trajectory and $dQ/dx$ are fitted again.
Segmenting the track helps prevent track fitting from bypassing a kink by requiring the trajectory to pass through the vertex connecting the segments. This process is repeated until no more kinks are found.
\item Track segments associated with existing track trajectories are removed from the TPC cluster and the remaining pieces are examined to search for new track segments.
The newly found track segments are connected back to the existing track segments to seed new vertices. Track trajectory and $dQ/dx$ fitting is repeated until no new track segments are found.
\item Additional empirical algorithms are used to merge very close vertices to each other. For more details, please see Ref.~\cite{code_merge_vertices}.
\end{itemize}
The order of examinations (e.g. searching for kinks along a track or perform new track fitting) is fixed given a certain input so that the results of applying the aforementioned algorithms are reproducible. 
Figure~\ref{fig:multi-track-fitting} shows an example of the multi-track trajectory and $dQ/dx$ fitting for the event shown in figure~\ref{fig:illustration_pattern_recognition}. 

\subsection{3D Vertex Fitting}~\label{sec:vertex_fit}
In order to improve the quality of the multi-track trajectory and $dQ/dx$ fitting (described in section~\ref{sec:multi_track_fitting}), a dedicated algorithm is introduced to better determine the position of the vertex when multiple track segments are connected. Note, this vertex fitting algorithm is used to improve the precision of all vertices, both primary and secondary. The neutrino vertex (primary vertex) identification algorithms introduced later in section~\ref{sec:nu_vtx} and~\ref{sec:dl_nu_vtx} are used to find the primary vertex.

\begin{figure}[thb]
  \centering
  \includegraphics[width=0.99\textwidth]{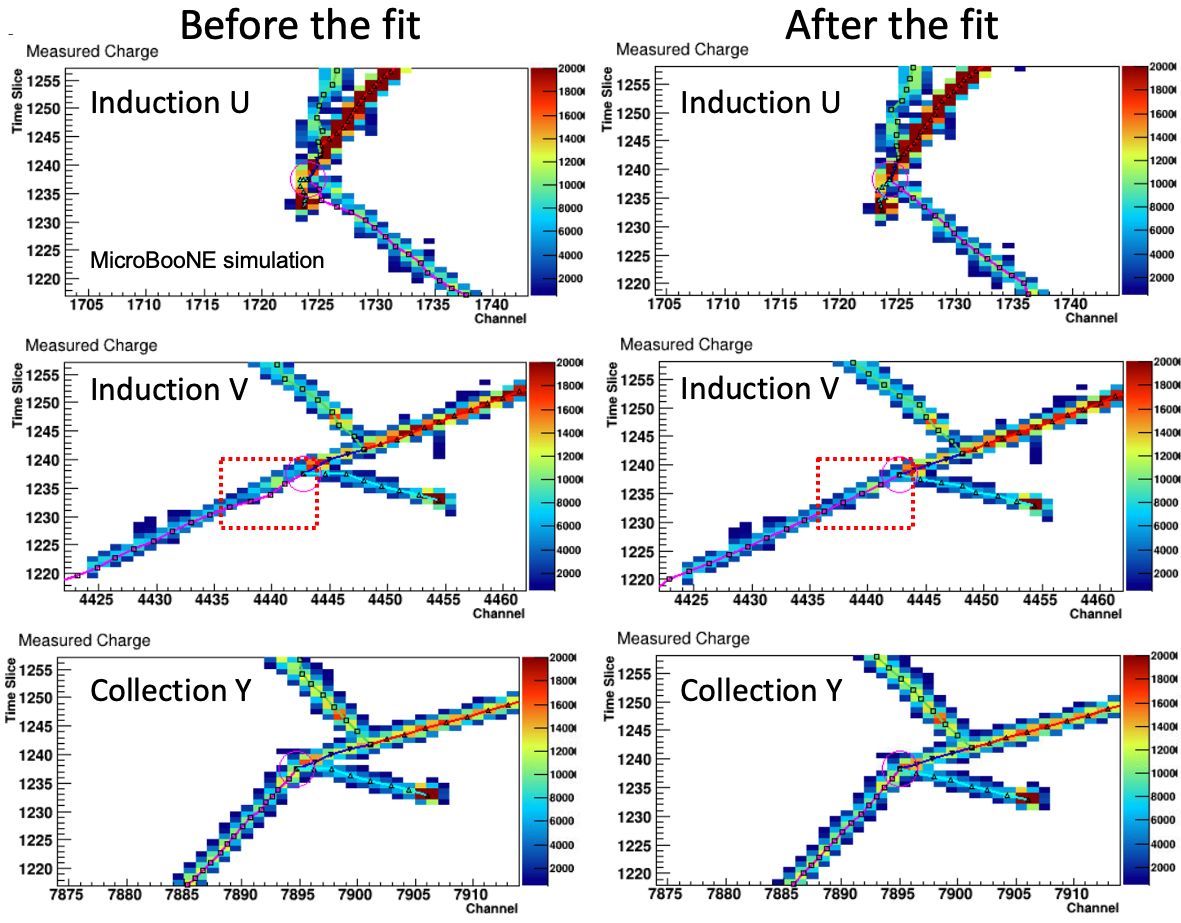}
  \put(-300, 335){MicroBooNE data (run 5384, event 2561)}
  \caption{ Best-determined track trajectories before (left) and after (right) the 3D 
  vertex fitting are displayed as lines on the reconstructed ionization charge.
    The $z$-axis represents the amount of ionization charge.
    The improvement of trajectory fitting near the vertex region (highlighted 
    with dashed red boxes) is most prominent in the middle row plots 
    (induction V plane). The original bias in the fitted track trajectory is 
    the result of the incorrect association of the 2D charge given that the initial 
    track trajectory is determined from the shortest path algorithm. The accurate
    vertex position determination allows the shortest path algorithm to provide
    improved initial track trajectories to determine the charge association for the 
    fitting algorithm.
  }
  \label{fig:vtx_fit}
\end{figure}

Instead of operating on 2D projections, the vertex fitting procedure is performed on the 3D space points.
First, we denote the position of the vertex to be determined as $\left(x,y,z\right)$. For the 3D space points associated with each track segment $i$, the points that are within a certain range (with a minimal distance of 1.5~cm and a maximal distance of 6~cm) are selected. Requiring a minimal and maximal distance reduces the bias in the fit, as points too close to the vertex can bias the fit due to incorrect associations with track segments and points too far from the vertex can bias the fit for tracks that change direction. A principle component analysis (PCA) is then performed:
\begin{itemize}
\item The center of the 3D points is $\left(x_{i0}, y_{i0}, z_{i0} \right)$. We further define $\vec{r}_i = \left( x_{i0}-x, y_{i0}-y, z_{i0}-z \right)$.  
\item The three eigenvectors of the PCA are denoted as $\vec{v}_{i1}$, $\vec{v}_{i2}$, and $\vec{v}_{i3}$, with the eigenvalues as $\lambda_{i1}$, $\lambda_{i2}$, and $\lambda_{i3}$, respectively. The distance from the vertex position to the main axis (first principal component) of the 3D points can thus be written as $\left( \vec{r}_{i} \cdot \vec{v}_{i2} \right)$ and $\left( \vec{r}_{i} \cdot \vec{v}_{i3} \right)$ for the second and third principal component directions respectively.
\end{itemize}
With these definitions, we can thus form a global test statistics to determine the vertex position:
\begin{equation}\label{eq:3d_vtx}
  T = \sum_i \left( \frac{\lambda_{i1}}{\lambda_{i2}} \left( \vec{r}_{i} \cdot \vec{v}_{i2}  \right)^2  + \frac{\lambda_{i1}}{\lambda_{i3}} \left( \vec{r}_{i} \cdot \vec{v}_{i3} \right)^2 \right) + \lambda \cdot \left(
  \left( x - x_{org}\right)^2 + \left( y - y_{org}\right)^2 + \left( z - z_{org}\right)^2  \right),
\end{equation}
where the first term in the summation minimizes normalized distance between the vertex and the main axis of the 3D points from a track segment. The second term in Eq.~\eqref{eq:3d_vtx} minimizes the distance between the vertex position and its initial value $\left( x_{org}, y_{org}, z_{org}\right)$ with a regularization strength $\lambda$. The addition of this term is to avoid unrealistic fitting results in certain situations (e.g. the track segments are all nearly parallel to each other). The minimization of Eq.~\ref{eq:3d_vtx} can be solved with linear algebra, which ensures the stability of the fitting procedure. Figure~\ref{fig:vtx_fit} shows the best-determined track trajectories before 
and after the 3D vertex fitting. The improvement close to the vertex is most prominent 
in the middle row plots.

\subsection{Separation of Track and Shower Topology}~\label{sec:track_shower_sep}
We use three methods to differentiate tracks from electromagnetic (EM) showers at different energies. They are: i) multiple Coulomb scattering measurements to differentiate track-like electrons from other particles at low energy, ii) the presence of additional isolated clusters in close proximity to track-like electrons, also at low energy, and iii) the EM shower width perpendicular to the trajectory direction for high energy electrons. We discuss each method in more detail below.

\begin{figure}[thb]
  \centering
  \includegraphics[width=0.7\textwidth]{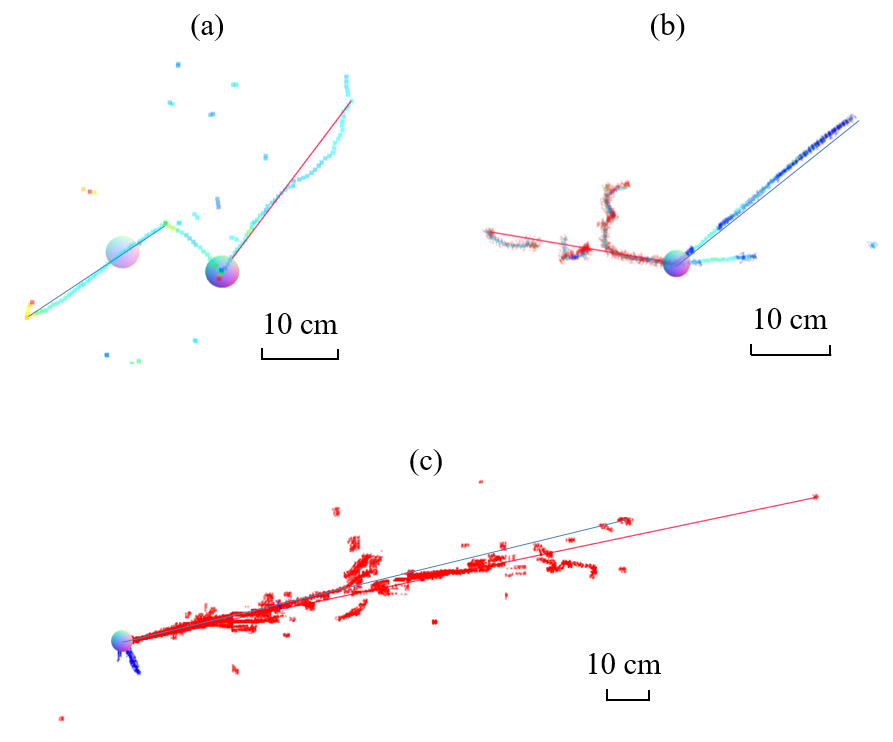}
  \put(-340, 208){\small MicroBooNE data}
  \put(-140, 208){\small MicroBooNE data}
  \put(-300, 60){\small MicroBooNE data}
  \caption{Example of various EM showers. (a) A very wiggled track. Color represents reconstructed charge along the fitted track trajectories. The blue (red) color indicates
  low (high) ionization charge. (b) A scattered track with two isolated showers aligned with a straight line. Shower-like objects are indicated by red color overlay; track-like objects are indicated by blue color overlay. (c) A high-energy EM shower. Shower-like objects are indicated by red color overlay; track-like objects are indicated by blue color overlay. Rainbow circles indicate particle starting positions in all three panels. 
  }
  \label{fig:track_shower}
\end{figure}

At low energies ($\mathcal{O}(\text{10 MeV})$), the bremsstrahlung effect becomes less important for electrons compared 
to the ionization effect. The low-energy electrons could behave like tracks instead of electromagnetic showers (i.e. multiple tracks including electron-positron pairs produced by the bremsstrahlung photons). Nevertheless,
the combination of an electron's small mass and ample interactions with atomic electrons in liquid argon means 
low-energy electrons have more large-angle scattering than other particles.
An optimal way to differentiate low-energy electrons from low-energy muons, pions, or protons in LArTPCs is to use the multiple Coulomb scattering (MCS) method.
For electrons, the MCS-derived momentum assuming a muon mass~\cite{MicroBooNE:2017tkp} will be much lower than that reconstructed from the track residual range. We use a simplified algorithm to differentiate a wiggled track from a straight track. We divide each track segment into many sections with each section about 10~cm long. For each track section, we calculate the straight-line distance between the starting and ending points. Along its fitted track trajectory, the trajectory length in this track section is also calculated. For a wiggled track segment, the direct length after summing over all track sections is expected to be much smaller than the trajectory length after summing over all sections. For a straight track segment, on the other hand, the ratio of these two lengths is close to one.~\footnote{Typically, a value of 0.9 or greater is required empirically.} Figure~\ref{fig:track_shower}a shows such an example.

For low-energy electrons, in addition to the above mentioned length ratio, the presence of nearby isolated clusters (i.e. bremsstrahlung photons emitted during the development of EM shower) can also be used to identify an EM shower. Figure~\ref{fig:track_shower}b shows an identified EM shower in which case two isolated clusters are associated with the wiggled track. The usage of isolated clusters for EM shower identification requires an alignment of the isolated clusters to the shower trunk (first few centimeters of the EM shower) direction. A good neutrino vertex reconstruction plays an important role in determining the shower trunk and its direction.

For high-energy electrons (see Fig.~\ref{fig:track_shower}c), the resulting EM showers tend to extend further in the projections perpendicular to the momentum direction because a large number of secondary electrons and positrons are produced. The measurement of the perpendicular spread of associated 3D space points (reconstructed in the 3D imaging step) along the EM shower direction can be used to identify an EM shower. The evolution of the width along the EM shower direction (e.g. a gradual increase of the width) can also be used to determine the starting point of the EM shower, which can be further used to determine the direction of the EM shower using the stem of the EM shower (track-like segment at the beginning of an EM shower). The direction of the EM shower is a crucial input for neutrino vertex reconstruction. For any projective readout, one expects ambiguities in the transverse plane (parallel to the readout plane). The measurement of widths must take this into account. We define the direction of the shower to be $\vec{r}_{mom}$ and the direction of drifting electrons (perpendicular to the transverse plane) to be $\vec{r}_{drift}$. The measurement of the widths is performed along the direction of $\vec{r}_{mom} \times \left(\vec{r}_{mom} \times \vec{r}_{drift}\right)$, which is perpendicular to both the momentum direction $\vec{r}_{mom}$ and the vector representing the ambiguities $\vec{r}_{mom} \times \vec{r}_{drift}$.

\subsection{Neutrino Vertex Identification with Traditional Techniques}~\label{sec:nu_vtx}
\begin{figure}[H]
  \centering
  \includegraphics[width=0.8\textwidth]{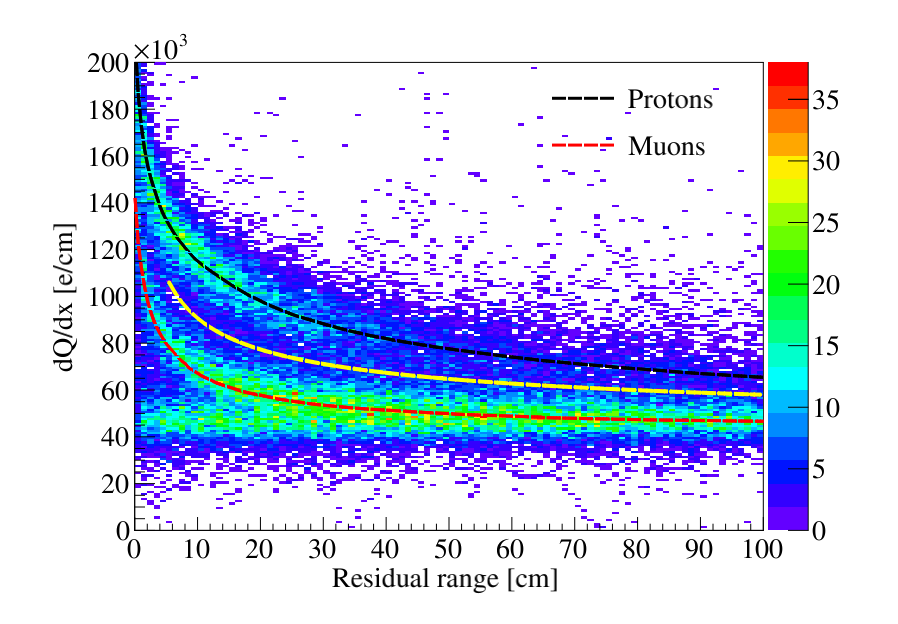}
  \put(-130, 218){MicroBooNE data}
  \caption{
    $dQ/dx$ vs residual range from MicroBooNE data. Two clearly separated bands, indicated by black and red lines, represent protons and muons. The measurement of $dQ/dx$ near the end point of the track can be used to determine the track direction and particle identification. The band around $dQ/dx\sim$ 50k e/cm extending to zero residual range represents the muons exiting the detector. The black and red curves are the predictions of most probable $dQ/dx$ values for protons and muons, respectively, considering the recombination effect as reported in Ref.~\cite{Adams:2019ssg}. The yellow curve indicates the empirical cut to separate protons and muons/pions.
  }
  \label{fig:pid}
\end{figure}

The reconstructed particle direction provides important information for the neutrino interaction vertex identification.  In the primary neutrino interaction vertex, the particles entering into the vertex would be
the neutrino (invisible to the detector) and the argon nuclei (also invisible to the detector). Therefore, 
all tracks and EM showers would travel outward from the primary neutrino interaction vertex. The situation
is different for a secondary interaction vertex in the main interaction cluster containing the true neutrino interaction vertex. In this case, the particles entering into the vertex would be a charged particle 
(e.g. produced by the primary neutrino interaction) and an argon nuclei (invisible to the detector). 
The rest of tracks and EM showers would travel outward from the secondary vertex. Given each interaction
always involves an argon nuclei, the chance to have two visible particle tracks entering into a vertex
is negligible.  Therefore, the determination of the track and EM shower directions is crucial. As described in section~\ref{sec:track_shower_sep}, the gradual increase of the EM shower's width can be used to determine its direction.
For tracks, the identification of a Bragg peak can be used to determine the direction.
Figure~\ref{fig:pid} shows the $dQ/dx$ distribution as a function of the residual range (distance to the end point of a track).
The $dQ/dx$ rise can be used to mark the end point of a track.
In addition, two clearly separated bands, indicated by black and red lines, correspond to protons and muons/pions.
And this can be used to separate protons from muons/pions.

In addition to the track and EM shower's directional information, the following information is taken into consideration in determining the neutrino vertex (primary vertex): 
\begin{itemize}
\item The position of the vertex. For example, the neutrino vertex in an interaction cluster tends to be more upstream than any other activity near the neutrino beam direction. On the other hand, a vertex at the detector boundary is less likely to be the neutrino vertex.
\item The number of directly connected tracks and EM showers. Statistically, more tracks and showers are connected to the neutrino vertex than secondary vertices.
\item A set of connectivity rules for the activities (tracks or showers) near the vertex. For example, there should not be two tracks going into the same vertex. There should not be one EM shower going into a vertex and a track coming out of the same vertex. In addition, the situation that one track comes in and another track goes backward with a large angle is not preferred. 
\item Some vertex candidates can be excluded by the local event topology. For example, vertices inside an EM shower and vertices from a delta-ray emerging from a straight muon track can be safely excluded. For details, please see Ref.~\cite{code_examine_main_vertex}.
\end{itemize}
In practice, each piece of information is summarized into a score.
The vertex candidate with the highest score is determined to be the primary neutrino vertex.
Given this method is surpassed by the Deep-learning neutrino vertex determination (discussed in Sec.~\ref{sec:dl_nu_vtx}), the technical details of the score evaluation are skipped.

\begin{figure}[thb]
  \centering
  \includegraphics[width=0.99\textwidth]{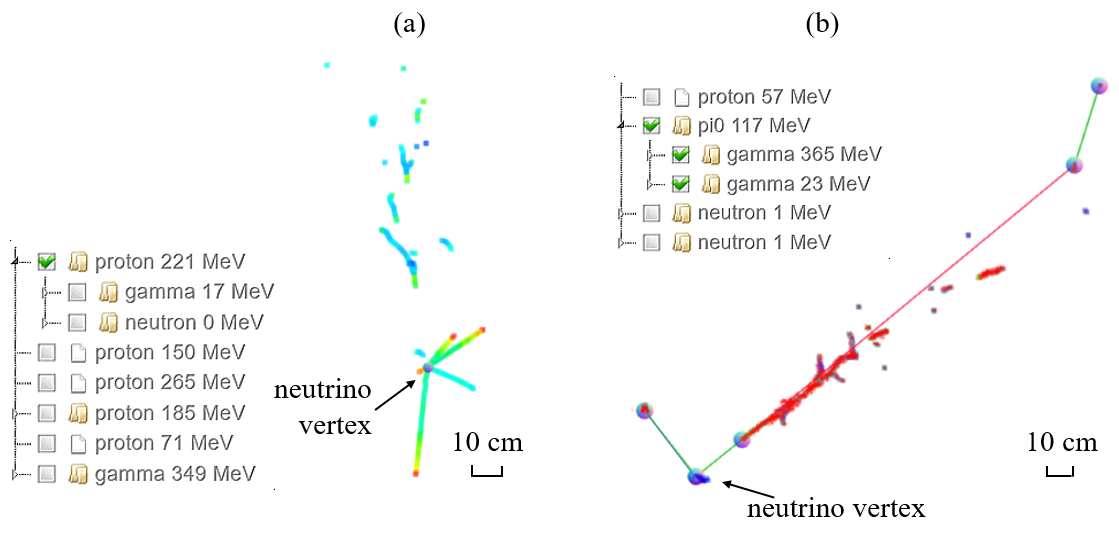}
  \put(-420, 175){MicroBooNE data}
  \put(-100, 175){MicroBooNE data}
  \caption{ Illustration of the primary neutrino vertex determination. (a) the neutrino vertex is determined to be inside the main TPC cluster consisting of multiple tracks. (b) the neutrino vertex is determined to be at a short proton where two separated EM showers point back toward. This is a candidate NC$\pi^0$ production event. The reconstructed particle flow information is also displayed (Refer to the caption of figure~\ref{fig:illustration_pattern_recognition} for more details).
  }
  \label{fig:nu_vtx}
\end{figure}

Using the strategies above, one neutrino vertex candidate can be identified in each TPC cluster. 
Given there could be multiple TPC clusters in each neutrino event, another algorithm is used to select the primary interaction vertex (neutrino vertex) from all vertex candidates, since the primary neutrino vertex may be outside the main TPC cluster (the TPC cluster most closely matched to the observed light pattern). Therefore, we utilize the position of each candidate vertex and the size and direction of its associated TPC cluster to select primary vertices among candidates in all TPC clusters. Figure~\ref{fig:nu_vtx}a shows an example case where the neutrino vertex is inside the main TPC cluster consisting of multiple tracks. Figure~\ref{fig:nu_vtx}b shows an example case where the neutrino vertex is located on a short proton track with two separated EM showers (reconstructed to be from a neutral pion decay) pointing back toward it.

The performance of the primary neutrino vertex identification with these traditional pattern recognition techniques has limitations. Among $\nu_\mu$CC events ($\nu_\mu$ charged-current interactions), about 70\% of neutrino vertices are correctly determined, where a reconstructed vertex position is defined to be correct if it lies within 1 cm of the true neutrino vertex position. Among $\nu_e$CC events ($\nu_e$ charged-current interactions), about 50\% of neutrino interaction vertices are correctly identified. We describe further improvements of the neutrino vertex identification using deep learning techniques in section~\ref{sec:dl_nu_vtx}.

\subsection{Neutrino Vertex Identification with a Regressional Segmentation Network}~\label{sec:dl_nu_vtx}
While traditional pattern recognition techniques provide reasonable performance in identifying the primary neutrino interaction vertex (section~\ref{sec:nu_vtx}), an improved performance is desired to enhance the selection efficiency of $\nu_e$ and $\nu_\mu$ charged-current interactions. We use a Deep Neural Network (DNN) on the reconstructed 3D space points and vertex candidates to further enhance the primary neutrino interaction vertex identification performance. 

This neutrino vertex identification problem can be considered as a special case of an object detection task in computer vision. Object detection usually takes 3 steps: region proposal, object type identification, and object segmentation~\cite{object_detection_dl}. In the vertex identification case, we are targeting only one type of object: the neutrino vertex and the region of the object, so we focus on the region proposal step. We use a customized regressional segmentation network to predict the distance map of each voxel (a 3D pixel) to the neutrino vertex, which improves the training efficiency compared to only predicting whether a voxel is a vertex or not. We use SparseConvNet~\cite{scn1,scn2}, which makes the model training and inferencing (the process of using a trained DNN model to make predictions against previously unseen data) much faster and more scalable.
Similar techniques were used in other MicroBooNE analyses~\cite{MicroBooNE:2020yze}.
The inferencing utilities are available in the Wire-Cell software package~\cite{wcp}.


\begin{figure}[H]
    \centering
    \includegraphics[width=0.99\figwidth]{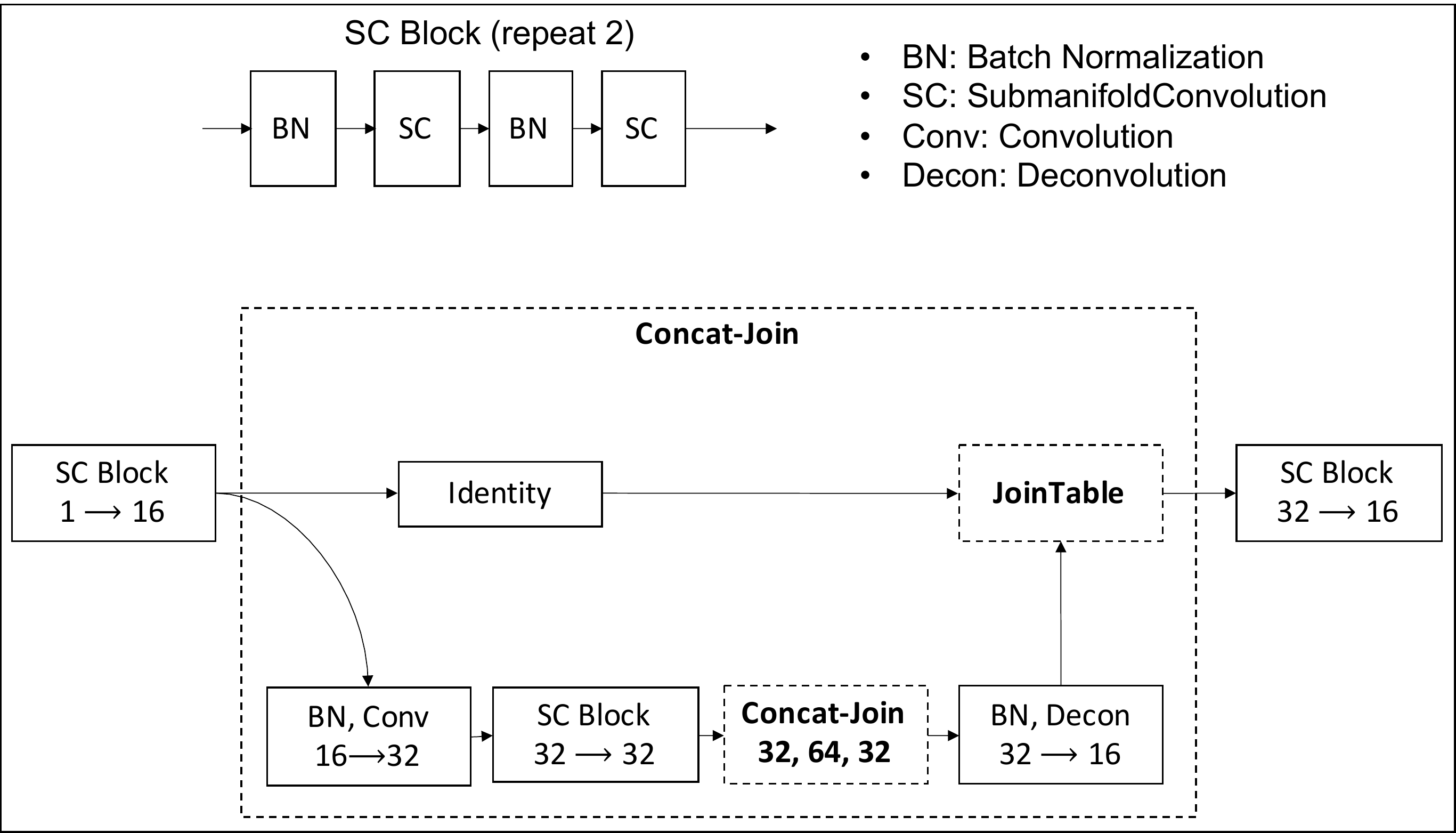}
    \caption{Illustration of the Sparse U-Net network structure from SparseConvNet~\cite{scn1,scn2}. 
    The bottom diagram shows a level 2 Sparse U-Net.}
    \label{fig:dl_vtx:net}
\end{figure}

\begin{figure}[H]
    \centering
    \includegraphics[width=0.99\figwidth]{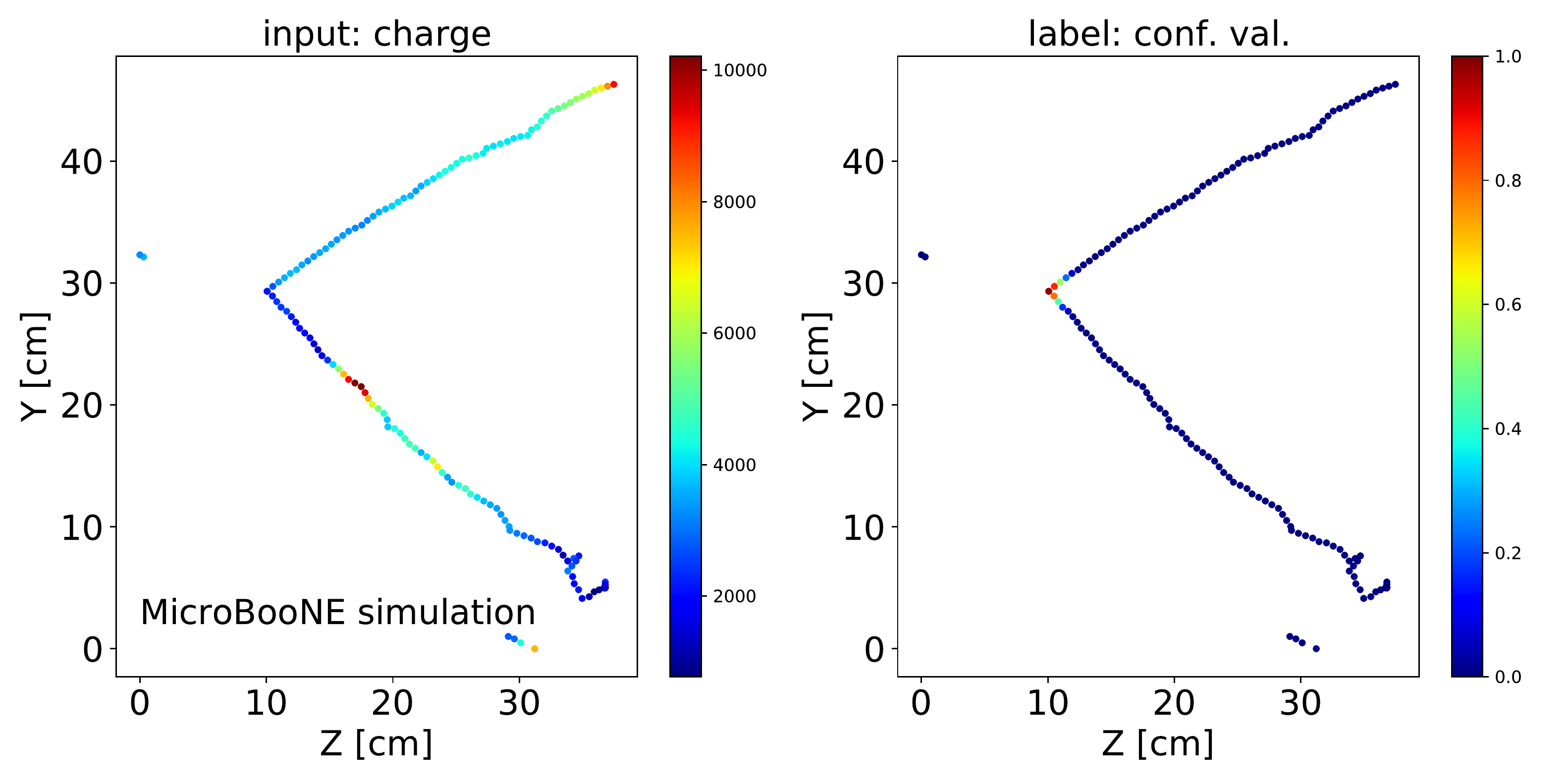}
    \caption{Example of input (charge) and label (confidence value) images before voxelization (showing a 2D projection). The color represents reconstructed charge in the left panel and confidence value in the right panel. From the right panel, we can see the confidence value becomes larger when getting closer to the vertex as defined in Eq.~\ref{eq:confval}.}
    \label{fig:dl_vtx:input-label-example}
\end{figure}

The key part of our regressional segmentation network is a Sparse U-Net~\cite{DBLP:journals/corr/RonnebergerFB15} which extracts a feature vector for each pixel or voxel.
Figure~\ref{fig:dl_vtx:net} shows details of a level 2 Sparse U-Net.
One can add more levels by iteratively inserting more ``Concat-Join'' blocks.
We use level 5 in this paper.
More implementation details can be found in~\cite{uboone-dl-vtx}.
The SparseConvNet works on sparsified 2D or 3D images, which consist of a list of pixels (2D) or voxels (3D). The Wire-Cell reconstructed 3D points are placed into voxels, which is a cube with 0.5~cm in each of three dimensions. If multiple 3D points fall into the same voxel, the values are averaged. The SparseConvNet takes two tensors as input. One is a coordinate tensor (integer type) with position information for each voxel. The other one is a feature tensor (float type) with the charge information. The output of the SparseConvNet is also a list of features for each voxel, for now we extract only one feature which we call \textit{confidence value} related to the distance between a point and the truth vertex. The truth label for this \textit{confidence value} is calculated before voxelization, and the equation used is:
\begin{equation}
    Conf_{\text{truth}} = \text{exp}\left( -\frac{\lVert \vec{x} - \vec{v}_\text{truth} \rVert ^{2}}{2\sigma^2}\right),
    \label{eq:confval}
\end{equation}
\noindent where $Conf_{\text{truth}}$ is the truth label of the \textit{confidence value}; $\vec{x}$ and $\vec{v}_\text{truth}$ are the reconstructed 3D charge and truth vertex positions; $\sigma$ is a regularization parameter (1~cm is used). \textit{Confidence value}s from multiple voxels (pixels) form a ``Confidence Map''~\cite{cao2018openpose}.
Figure~\ref{fig:dl_vtx:input-label-example} shows an example of the input (charge) and label before voxelization.

The dataset used for training the network consists of 48k $\nu_e$CC events simulated through the MicroBooNE simulation chain and Wire-Cell reconstruction. The dataset for validating the network performance consists of 4k $\nu_e$CC events. The dataset for testing the network performance consists of 4k $\nu_e$CC and 4k $\nu_\mu$CC events.
The total dataset size is chosen based on the available official MicroBooNE simulation production sample size.
Then we try to maximize the training set size while keeping enough for validation and testing.
The choice of focusing on the $\nu_e$CC event topology is motivated by the need to improve $\nu_e$CC event selection, which is much more challenging than the $\nu_\mu$CC event selection. Nevertheless, the final testing results of the network show an improvement over the performance of the traditional algorithm (described in section~\ref{sec:nu_vtx}) in both $\nu_e$CC and $\nu_\mu$CC events.

\begin{figure}[H]
    \centering
    \includegraphics[width=0.49\figwidth]{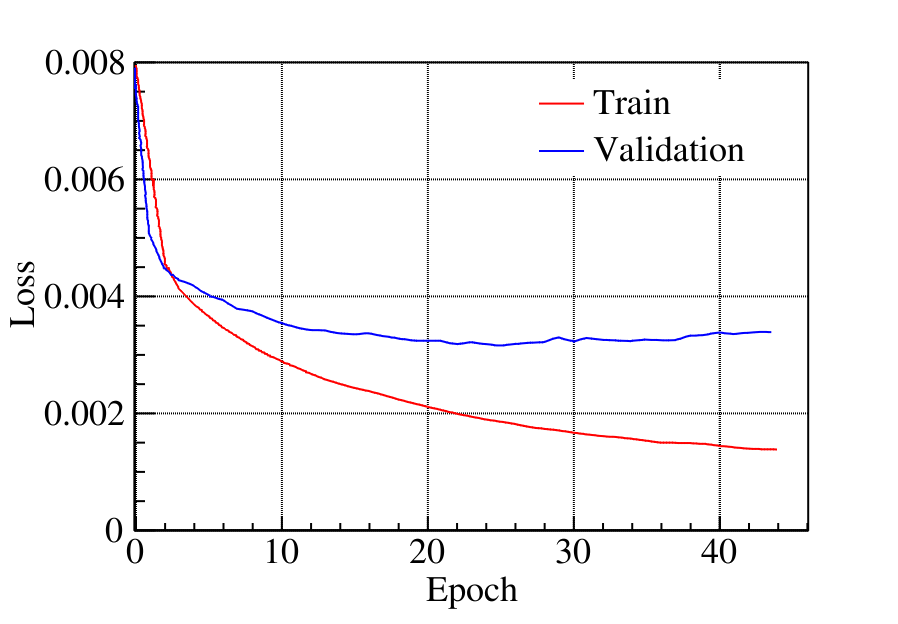}
    \put(-122, 128){\small MicroBooNE simulation}
    \includegraphics[width=0.49\figwidth]{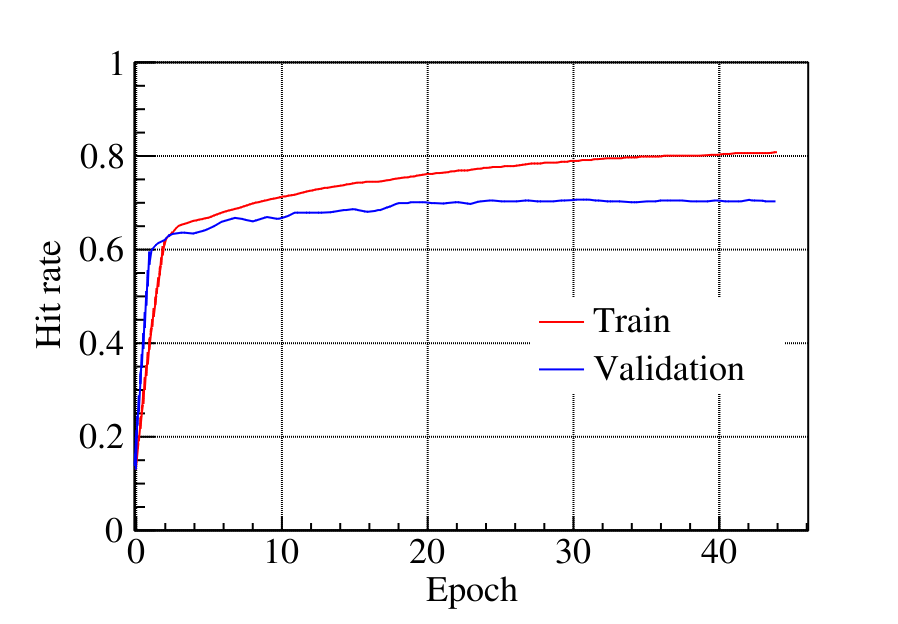}
    \put(-122, 128){\small MicroBooNE simulation}
    \caption{Left: epoch average loss as a function of epoch number for training (red) and validation (blue) samples; right: probability that the best predicted voxel is within 1 cm of the best truth voxel (voxel closest to the truth vertex).}
    \label{fig:dl_vtx:loss-epoch}
\end{figure}

\noindent Machine learning model parameters are estimated with the help of loss functions~\cite{python-ml}. The loss function we used is:
\begin{equation}
     \mathcal{L} = \frac{1}{N_{voxel}} \sum^{N_{voxel}} \left \| Conf_{pred} - Conf_{truth} \right \| ^2.
\end{equation}
The optimizer used is Adam~\cite{kingma2014adam}; the learning rate decays after each epoch as $l_0 \cdot exp(-n\cdot l_d)$, where $l_0$ is $10^{-5}$ and $l_d$ is 0.05 and $n$ is the epoch number. Figure~\ref{fig:dl_vtx:loss-epoch} shows the training progress, epoch average loss, and hit rate as a function of epoch number for the $\nu_e$CC training and validation samples. ``Hit rate'' here is defined as the probability that the best predicted voxel is within 1 cm of the truth voxel (the voxel closest to the truth vertex). From figure~\ref{fig:dl_vtx:loss-epoch}, we can see that:
i) the solution phase space is not fully covered by the training sample according to the gaps between the train and validation curves (we also observed the gaps getting smaller when increasing training samples);
ii) there are signs of over-training;
and iii) validation performance stopped improving after epoch 24.
The results shown in this paper are based on the model at epoch 24.

\begin{figure}[H]
    \centering
    \includegraphics[width=0.7\figwidth]{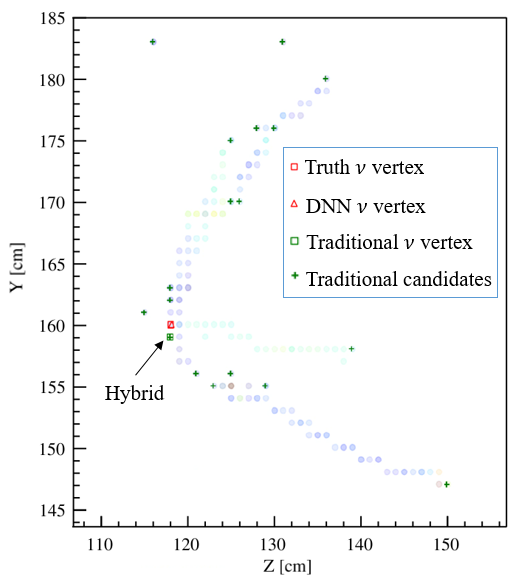}
    \put(-115, 316){MicroBooNE simulation}
    \caption{An event display to demonstrate finding the hybrid vertex based on a DNN vertex. Each filled circle represents a voxelized 3D charge projected on the Y-Z plane. In this example, the hybrid vertex (indicated by the black arrow) is different from the DNN one, which is better in this case.}
    \label{fig:dl_vtx:hybrid}
\end{figure}

\begin{figure}[H]
    \centering
    \includegraphics[width=0.8\figwidth]{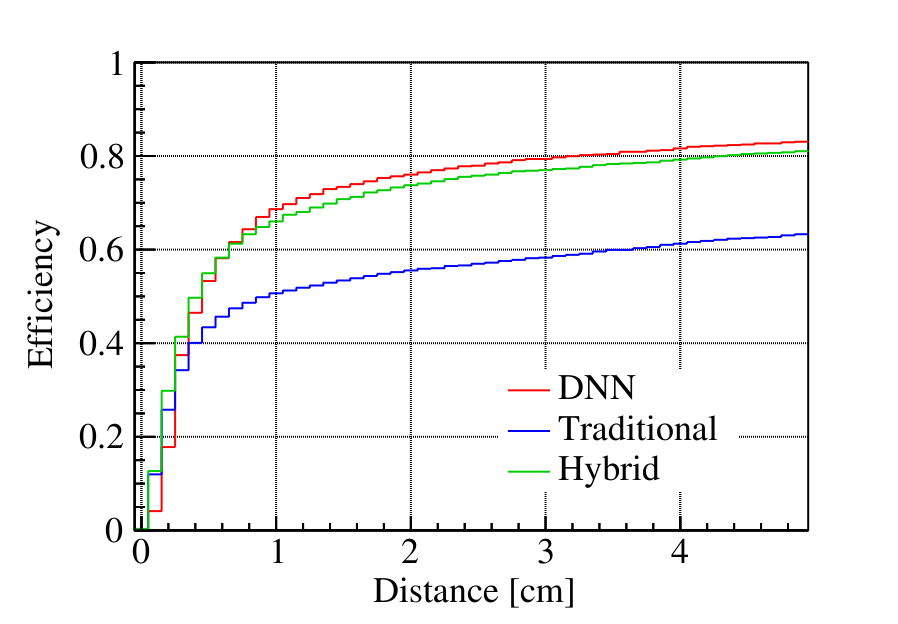}
    \put(-140, 208){MicroBooNE simulation}
    \caption{$\nu_e$CC neutrino vertex identification efficiency as a function of the maximum distance between the reconstructed and the truth, i.e., the numerator of the efficiency is defined as the number of events with a neutrino vertex reconstructed within $X$~cm of the truth, where $X$ is the horizontal coordinate of the plot. Three algorithms are compared: Traditional, DNN and Hybrid.
    }
    \label{fig:dl_vtx:nue-eff}
\end{figure}

\begin{figure}[H]
    \centering
    \includegraphics[width=0.8\figwidth]{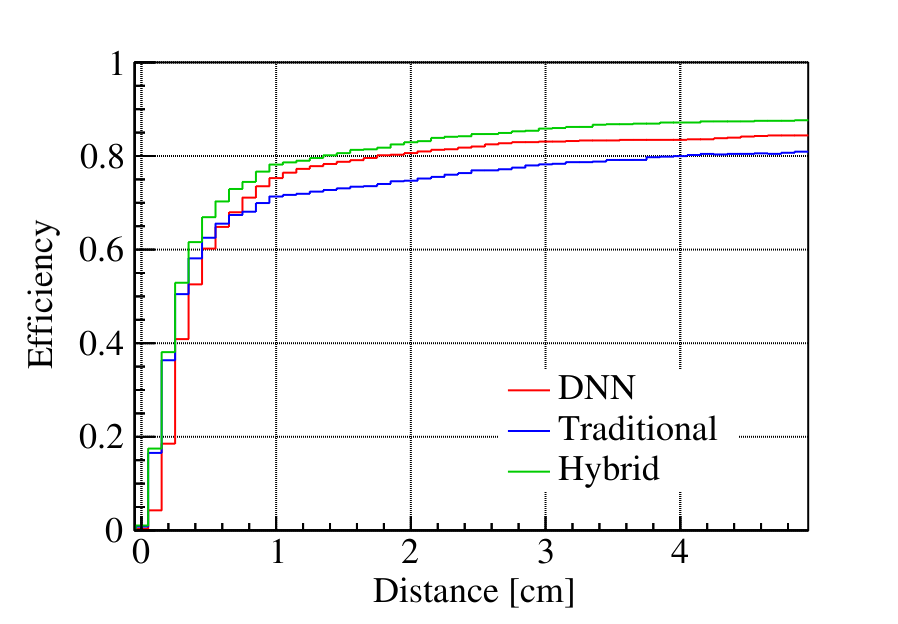}
    \put(-140, 208){MicroBooNE simulation}
    \caption{$\nu_\mu$CC neutrino vertex identification efficiency as a function of the maximum distance between the reconstructed and the truth, i.e., the numerator of the efficiency is defined as the number of events with a neutrino vertex reconstructed within $X$~cm of the truth, where $X$ is the horizontal coordinate of the plot. Three algorithms are compared: Traditional, DNN and Hybrid.}
    \label{fig:dl_vtx:numu-eff}
\end{figure}

To accommodate existing Wire-Cell reconstruction/pattern recognition algorithms, we select the neutrino vertex from the candidates provided by the traditional pattern recognition algorithms. We call this vertex the hybrid vertex. Figure~\ref{fig:dl_vtx:hybrid} shows an event display to demonstrate finding the hybrid vertex based on a DNN vertex. The decision proceeds as follows:
\begin{itemize}
    \item Find the voxel with the highest \textit{confidence value}, which we call the ``DNN Vertex''.
    \item Convert from voxel index back to continuous space coordinate.
    \item From the list of traditional vertex candidates, find the closest one to this DNN vertex in the coordinate space. (As described in section~\ref{sec:nu_vtx}, the traditional neutrino vertex is also selected from these traditional candidates.)
    \item Check if the distance between this traditional candidate and the DNN vertex is smaller than a threshold, currently 2~cm. Note this ``2~cm'' threshold used in the DNN-traditional vertex matching should not be confused with the ``1~cm'' reco-truth matching threshold used in the vertex finding algorithm performance evaluations.
    \item If the previous distance check is good ( $<$ 2 cm) use this traditional candidate as the ``hybrid vertex''; if not, use the ``reconstructed traditional neutrino vertex'' (section~\ref{sec:nu_vtx}) as the ``hybrid vertex''.
\end{itemize}
Figure~\ref{fig:dl_vtx:nue-eff} shows the performance of the trained model on the $\nu_e$CC test samples. Compared to the performance of traditional pattern recognition algorithm, the hybrid algorithm leads to a relative 30\% improvement in identifying the neutrino vertex.
At low distance cut ($\sim$0.5~cm) between the truth and the reconstructed neutrino vertex, the traditional algorithm outperforms the DNN algorithm, because of the finite voxel size (0.5~cm in each dimension) and the smearing of the reconstructed 3D space points. Figure~\ref{fig:dl_vtx:numu-eff} shows the performance of the current best model on the $\nu_\mu$CC test samples. The hybrid algorithm leads to a relative 10\% improvement in identifying the neutrino vertex, which is achieved on the $\nu_\mu$CC samples not used in the training. Furthermore, while the performance of the hybrid algorithm is slightly worse than that of DNN for $\nu_e$CC events, it is slightly better for $\nu_\mu$CC. This is because the performance of traditional algorithm for $\nu_e$CC is worse than that for $\nu_\mu$CC. As mentioned before, the hybrid vertex is used as the main result in the current workflow to accommodate existing Wire-Cell reconstruction/pattern recognition algorithms.
In the future, it is likely we will revise the downstream reconstructions directly based on the DNN vertexing results then we do not need to use the ``hybrid'' one.

\subsection{Evaluation of the Neutrino Vertex Identification Performance}\label{sec:eval_vtx}



The efficiency of the neutrino vertex identification plays a crucial role in $\nu_\mu$CC and $\nu_e$CC event selections. Therefore, we performed evaluations of the neutrino, primary muon, and primary EM shower vertex reconstruction.
{\it Pandora}~\cite{Acciarri:2017hat} is another state-of-the-art traditional pattern recognition technique\footnote{Some new developments of Pandora reconstruction start to involve deep-learning algorithms, but such developments were not included in this comparison and are yet to be used in the MicroBooNE experiment.} that is also used in MicroBooNE.
We compare the performance of the Wire-Cell vertex identification with Pandora as a benchmark. 

Two types of overlay MC samples are used to evaluate the neutrino vertexing performance. BNB $\nu$ overlay MC, which simulates neutrino interactions based on the BNB neutrino flux of which 99.5\% are $\nu_{\mu}$'s, is used to evaluate the performance for $\nu_{\mu}$CC and NC events. BNB intrinsic $\nu_e$ overlay MC, which is a $\nu_e$-enriched MC sample only simulating the electron neutrino interactions based on the $\nu_e/\bar{\nu}_{e}$ component in the BNB neutrino flux, is used to evaluate the performance for $\nu_{e}$CC events. 

\begin{figure}[!hbt]
		\begin{center}
\begin{overpic}[width=0.49\columnwidth]{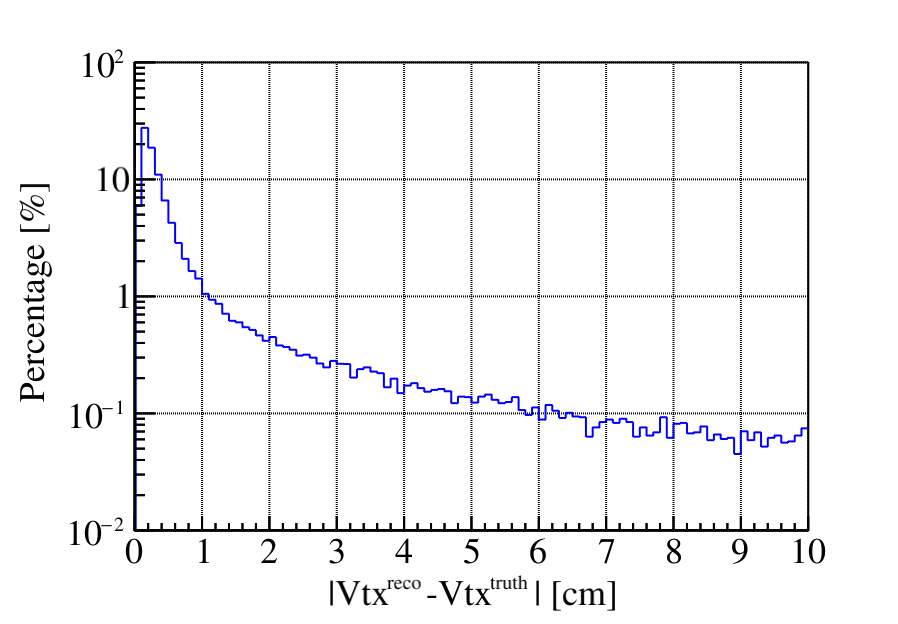}
   \put(42,64){\text{\small MicroBooNE simulation}}
\end{overpic}  
\begin{overpic}[width=0.49\columnwidth]{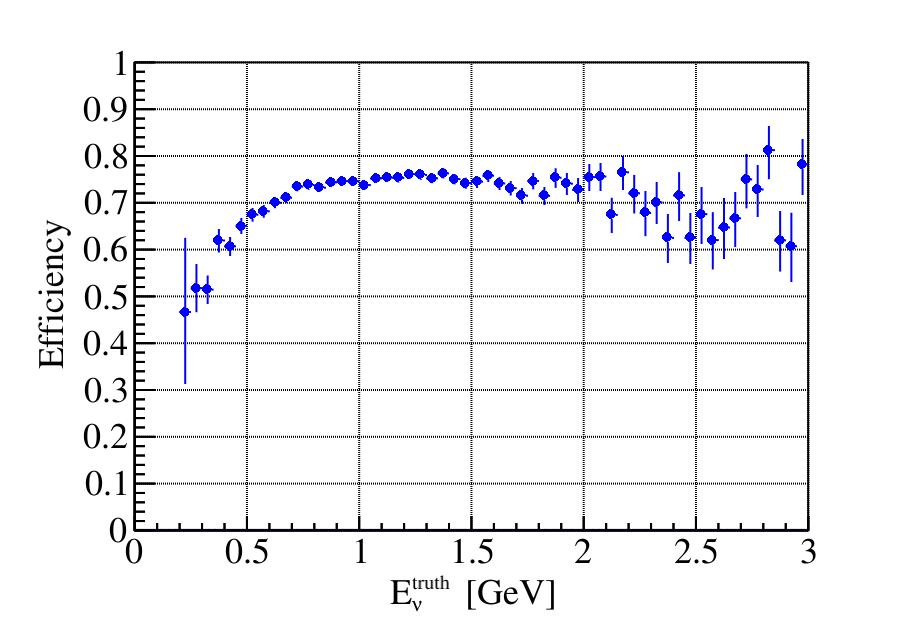}
   \put(42,64){\text{\small MicroBooNE simulation}}
\end{overpic}  
		\caption{Neutrino vertex reconstruction of the $\nu$ overlay sample, charged-current interaction, reconstructed with Wire-Cell. Left: distance between the reconstructed and truth neutrino vertex; Right: efficiency for events with vertex reconstruction position within 1~cm of the truth, as a function of truth neutrino energy. The error bars represent the statistical uncertainties.}
		\label{fig:nuVtxRes_CC}
		\end{center}
\end{figure}

\begin{figure}[!hbt]
		\begin{center}
\begin{overpic}[width=0.49\columnwidth]{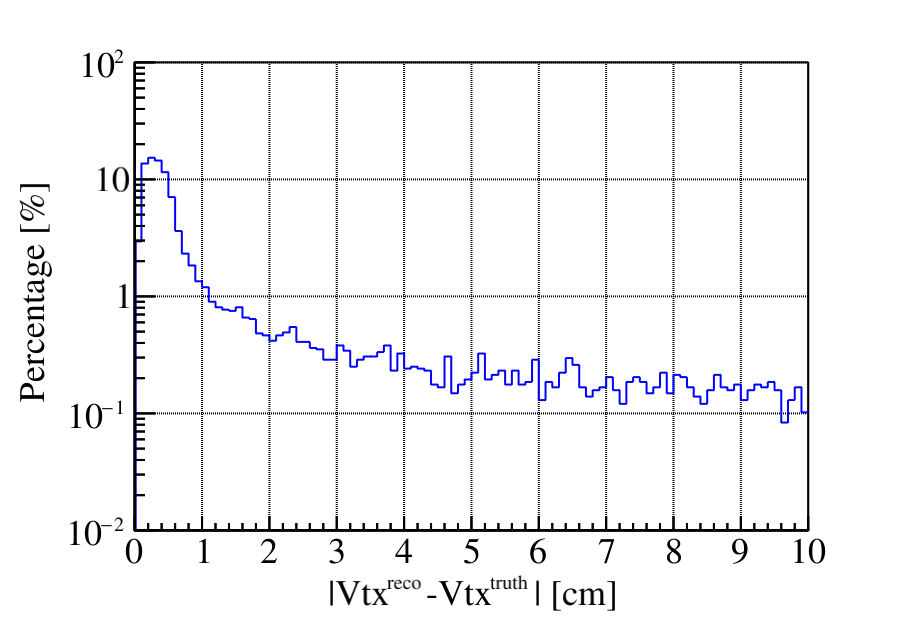}
   \put(42,64){\text{\small MicroBooNE simulation}}
\end{overpic}  
\begin{overpic}[width=0.49\columnwidth]{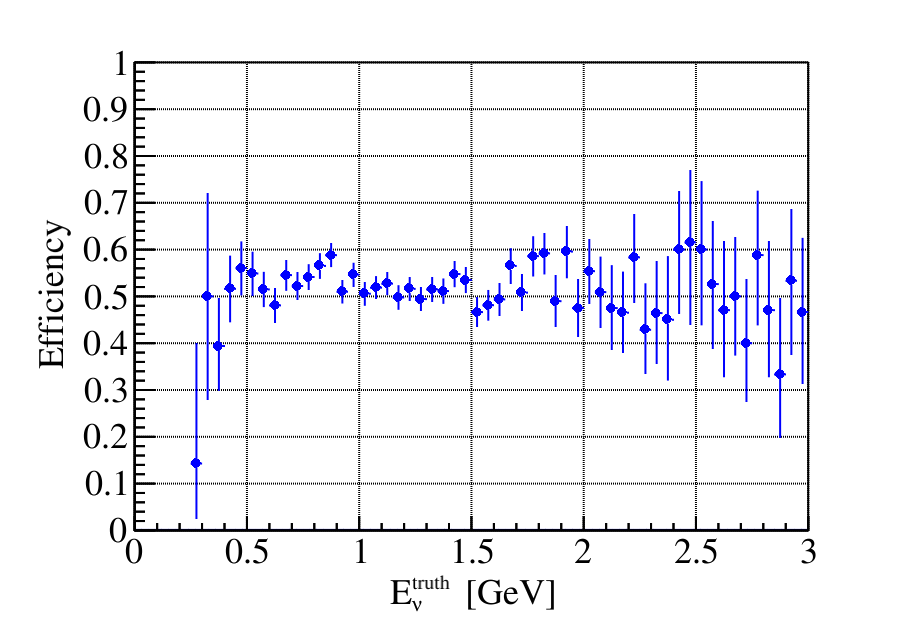}
   \put(42,64){\text{\small MicroBooNE simulation}}
\end{overpic}  
		\caption{Neutrino vertex reconstruction of the $\nu$ overlay sample, neutral-current interaction, reconstructed with Wire-Cell. Left: distance between the reconstructed and truth neutrino vertex; Right: efficiency for events with vertex reconstruction position within 1~cm of the truth, as a function of truth neutrino energy. The error bars represent the statistical uncertainties.}
		\label{fig:nuVtxRes_NC}
		\end{center}
\end{figure}

\begin{figure}[!hbt]
		\begin{center}
\begin{overpic}[width=0.49\columnwidth]{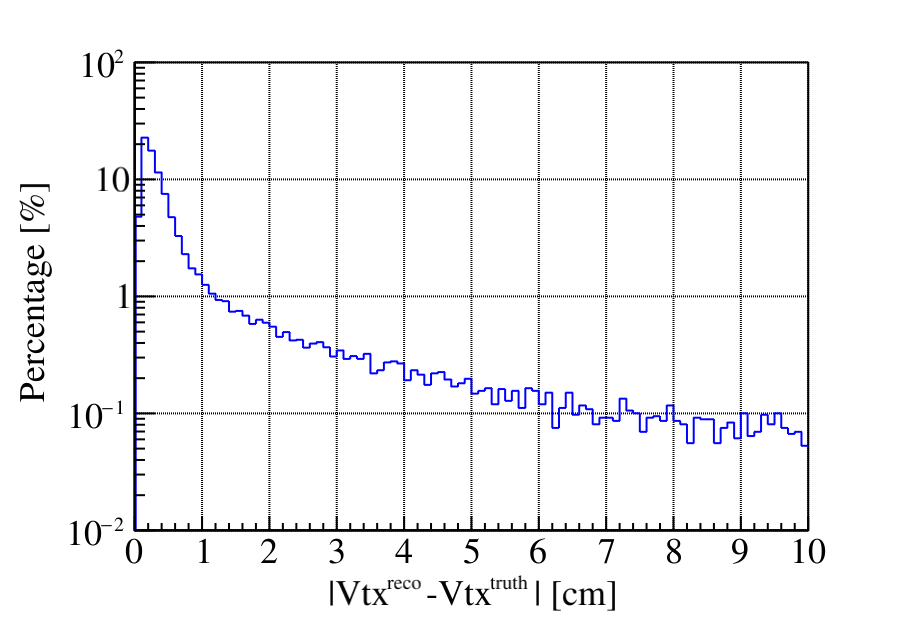}
   \put(42,64){\text{\small MicroBooNE simulation}}
\end{overpic}  
\begin{overpic}[width=0.49\columnwidth]{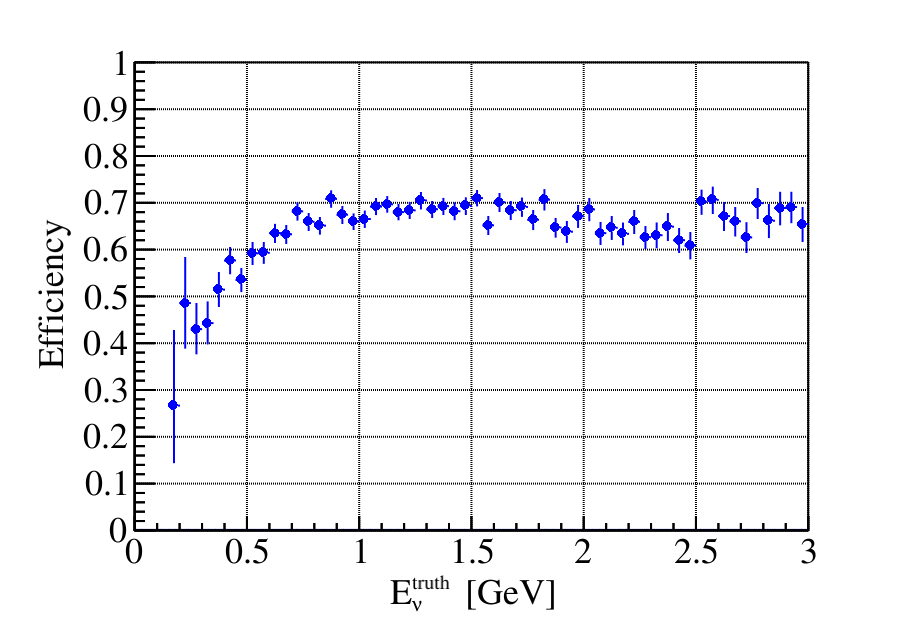}
   \put(42,64){\text{\small MicroBooNE simulation}}
\end{overpic}  
		\caption{Neutrino vertex reconstruction of the intrinsic \nue\ overlay sample, charged-current interaction, reconstructed with Wire-Cell. Left: distance between the reconstructed and truth neutrino vertex; Right: efficiency for events with vertex reconstruction position within 1~cm of the truth, as a function of truth neutrino energy. The error bars represent the statistical uncertainties.}
		\label{fig:nuVtxRes_nue}
		\end{center}
\end{figure}

The left panels of figure~\ref{fig:nuVtxRes_CC}, \ref{fig:nuVtxRes_NC}, and \ref{fig:nuVtxRes_nue} show distributions of the distance between the reconstructed neutrino vertex and true neutrino vertex for charged-current events in BNB $\nu$ overlay, neutral-current events in BNB $\nu$ overlay, and charged-current events in BNB intrinsic \nue\ overlay samples, respectively.
Percentages of events with a neutrino vertex reconstructed within 1~cm of the truth for these three samples are 70.3\%, 49.3\% and 65.2\%; within within 5~cm are 84.8\%, 63.5\% and 80.0\%, respectively.
In the right panels of these figures we calculate the percentage of events with the reco-truth vertex distance less than 1~cm.

\begin{table*}[!hbt]
	\centering
	\caption{The performance with the distance between reconstructed neutrino vertex and true neutrino vertex less than 1 cm. The performance are estimated with a BNB $\nu$ overlay sample, while charged-current \nue\ event efficiency are estimated with a BNB intrinsic \nue\ overlay sample.
	}
	\begin{tabular}{cccc}
		\hline
		\hline
		& Pandora reconstruction & Wire-Cell Traditional & Wire-Cell DNN Hybrid\\
		\hline
		Charged-current $\nu_{\mu}$& 70.3\% & 67.1\% &72.9\%\\
		\hline
		Neutral-current & 49.3\% & 50.6\% &52.1\%\\
		\hline
		Charged-current \nue\ & 65.2\% & 50.1\% &65.8\%\\
		\hline
		\hline
	\end{tabular}
	\label{tab:vtxRes_tab}
\end{table*}

Table~\ref{tab:vtxRes_tab} summarizes the performance comparison between Wire-Cell and Pandora~\cite{Acciarri:2017hat} reconstructions.
In order to make a fair performance comparison focusing on reconstruction algorithms, the same input events, with cosmic-rays cleaned up and unresponsive detector areas recovered using the upstream Wire-Cell imaging and clustering algorithms~\cite{Abratenko:2020hpp}, are feed into both Wire-Cell and Pandora reconstruction workflows.
Except for the inputs, the Pandora configuration is the same as used in Ref.~\cite{PeLEE_PRD}.
For charged-current events in the BNB $\nu$ overlay sample, Pandora reconstruction results in 70.3\% good reconstructed vertices (within 1~cm of the true vertex), and Wire-Cell input plus Wire-Cell reconstruction leads to 67.1\% (without DNN Vertexing) and 72.9\% (with DNN Vertexing) good reconstructed vertices.
For the neutral-current case, the Pandora reconstruction has 49.3\% efficiency while the Wire-Cell reconstruction has 50.6\% (without DNN Vertexing) and 52.1\% (with DNN Vertexing).
Neutrino vertex identification is generally harder for neutral-current interactions due to the absence of primary charged particles connecting to the neutrino interaction vertex and more complicated event topology.
When the BNB intrinsic \nue\ sample is used, Pandora has 65.2\% efficiency while Wire-Cell has 50.1\% (without DNN Vertexing) and 65.8\% (with DNN Vertexing) efficiency. The DNN Vertexing model is trained on a \nue\ charged-current sample and boosts the Wire-Cell vertexing performance by about 30\% for samples of the same type while it does not degrade the performance on samples of other types. After using the DNN Vertexing, the Wire-Cell vertexing performance is comparable with Pandora reconstruction for all samples tested.

\begin{figure}[!ht]
		\begin{center}
		\includegraphics[width=\columnwidth]{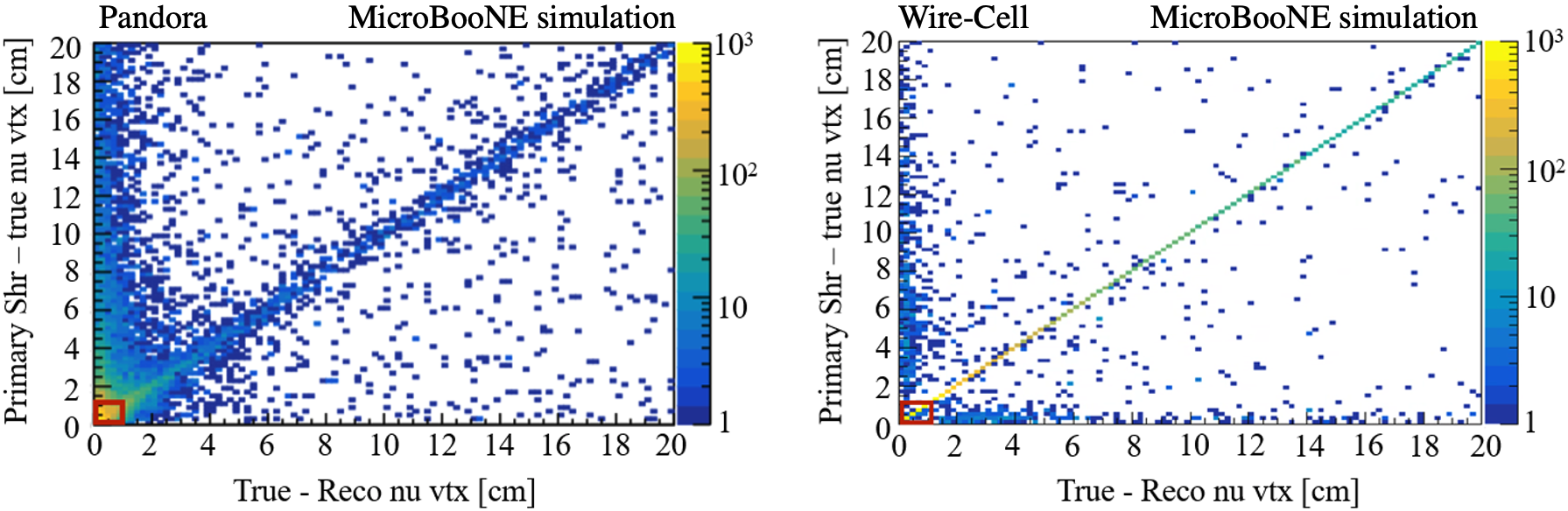}
		\caption{2D distribution of reconstructed primary shower vertex-true neutrino vertex difference vs. reconstructed neutrino vertex-true neutrino vertex of BNB intrinsic \nue\ sample, with reconstructed with Pandora (left) and with Wire-Cell (right). The two 2D histograms contain the same number of events. For Wire-Cell, the distribution is more concentrated along the diagonal line, which shows the reconstructed neutrino vertices and EM shower vertices are better aligned. Red boxes in the lower left corner indicates events with the position difference less than 1~cm. Note that the z-axis is in log scale.}
		\label{fig:nuVtxRes_2D}
		\end{center}
\end{figure}

As noted earlier, the BNB intrinsic \nue\ events can be used for the shower vertex reconstruction study as there are only $\nu_e$CC interactions in these events, which are supposed to have an EM shower attached to each neutrino vertex. 
Figure~\ref{fig:nuVtxRes_2D} is a 2D distribution of distance between the reconstructed primary shower vertex and the true neutrino vertex vs the distance between the reconstructed neutrino vertex and the true neutrino vertex.
For the Pandora reconstruction, the percentage of events that have both the reconstructed primary shower vertex and reconstructed neutrino vertex within 1~cm of the true neutrino vertex is 30.8\%, while the Wire-Cell reconstruction achieves 62.0\%.
This is because that the hybrid Wire-Cell neutrino vertex is selected from a list of candidates including the shower vertices. So for $\nu_e$CC interactions, when the neutrino vertex is identified correctly, there is a high chance that it is exactly the same position of a shower vertex.
While the Pandora reconstruction takes a different approach that the shower vertices are reconstructed more independently from the neutrino vertex, so that the analyzers would have additional information to use.
The better aligned neutrino and shower vertex should enhance the capability to differentiate primary electrons from photons and provide better a better capability in identifying the gap between the photon EM shower vertex and the neutrino vertex.
Furthermore, compared to the 65.8\% efficiency in correctly identifying neutrino vertex, the lower efficiency number when requiring both a reconstructed primary shower vertex and a reconstructed neutrino vertex within 1~cm of the true neutrino vertex essentially gives a measure of the reconstruction efficiency for the primary EM shower.


In this section, we describe various pattern recognition techniques towards neutrino vertex identification. 65.8\% and 72.9\%  efficiencies have been achieved in the Wire-Cell reconstruction
for charged-current $\nu_e$ and $\nu_\mu$ interactions respectively for the MicroBooNE experiment.
\section{Particle Flow Tree Reconstruction after Neutrino Vertex Identification}~\label{sec:pr_from_vertex}
In the first iteration of the Wire-Cell pattern recognition (section~\ref{sec:pr_to_vertex}), we identified the neutrino vertex as well as track segments and EM shower segments.
In this section, we describe the reconstruction of complete EM showers and the particle flow tree including $\pi^{0}$ reconstruction.

\subsection{Overall Electromagnetic Shower Clustering}~\label{sec:em_clustering}
As explained in section~\ref{sec:introduction} and in ref.~\cite{Abratenko:2021bzb}, ``a TPC cluster'', which consists of TPC activities that are connected, is used to organize (or divide) 3D and 2D data.
Within a TPC cluster, track segments are identified as outlined in section~\ref{sec:multi_track_fitting}.
Such a data structure is not sufficient to describe electromagnetic (EM) showers, which typically consist of many separated TPC clusters. Therefore, once the primary neutrino vertex is identified (section~\ref{sec:nu_vtx}), a dedicated EM shower clustering algorithm is applied.
The clustering algorithm considers directional and distance information.
First, the EM shower candidates within the TPC cluster that contains the primary neutrino vertex are examined.
Second, along the direction of each EM shower candidate, the separated pieces within a certain distance (80~cm) and a certain angle (15 degrees) are clustered together.
Finally, isolated EM shower candidates are examined according to their distances to the primary neutrino vertex and the same EM shower clustering algorithm is performed.

\begin{figure}[thb]
  \centering
  \includegraphics[width=0.99\textwidth]{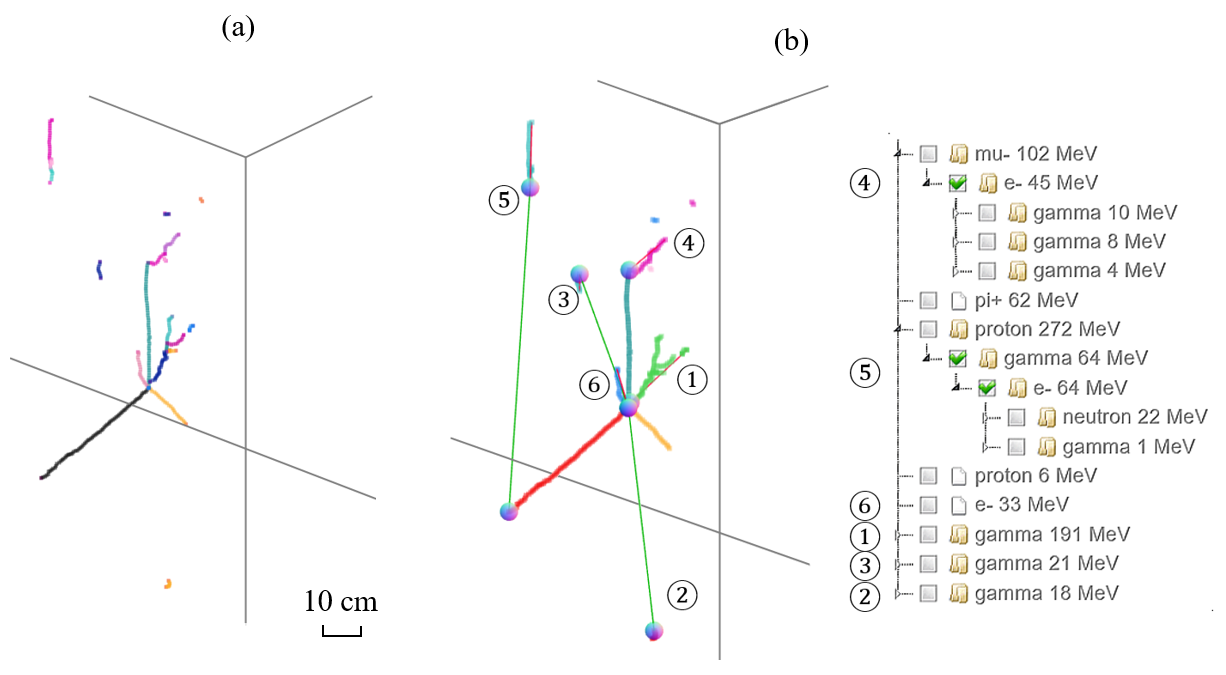}
  \put(-90,210){MicroBooNE data}
  \caption{Illustration of EM shower clustering. (a) shows the identified track segments with different colors in different TPC clusters. (b) shows the EM showers after clustering. The green lines illustrate the associations. The red lines connect the start and end points of each EM shower. The numbers of EM showers are also displayed with the particle flow information (Refer to the caption of figure~\ref{fig:illustration_pattern_recognition} for more details). }
  \label{fig:em_clustering}
\end{figure}

Figure~\ref{fig:em_clustering} shows an example of the EM clustering algorithm. Here, the first, second, third, and fifth EM showers are isolated EM showers.
The fourth EM shower is a Michel electron at the end of a stopped muon.
The sixth EM shower is connected to the primary neutrino vertex. It is
expected to be the result of a low-energy stopped charged pion, which is invisible because of the
detector spacial resolution. 

\subsection{Neutral Pion Reconstruction}~\label{sec:pio_reco}
With EM showers defined, one can reconstruct neutral pions which decay to two photons with a $\sim$99\% branching ratio. The reconstruction of neutral pions is useful in validating the energy scale for the EM shower energy reconstruction and is crucial for reconstructing the neutrino energy and in selecting events with neutral pions in both charged-current and neutral-current channels, which can be used to constrain backgrounds in single EM shower searches, for example the $\nu_e$CC selection or a single photon selection.

\begin{figure}[H]
  \centering
  \includegraphics[width=0.99\textwidth]{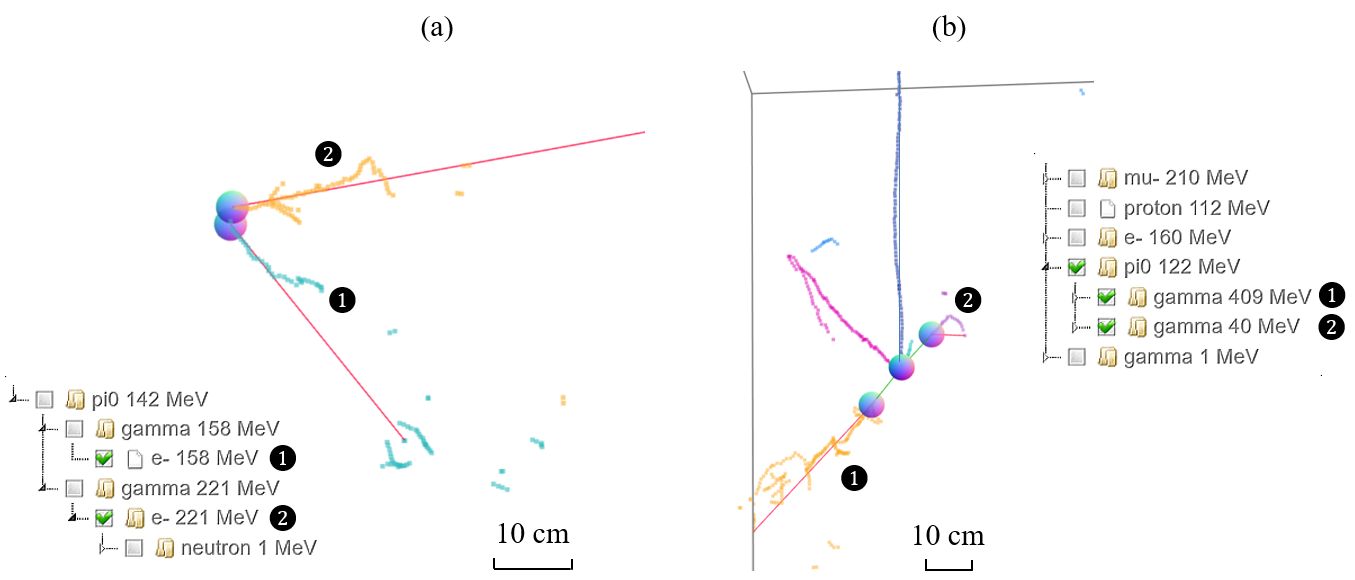}
  \put(-90,160){MicroBooNE data}
  \put(-425,160){MicroBooNE data}
  \caption{Examples of reconstructed neutral pions. (a): a neutral pion is reconstructed at an invariant mass of 142 MeV/c$^2$ without any TPC activity at its generation location. (b): a neutral pion is reconstructed at an invariant mass of 122 MeV/c$^2$ from the primary neutrino vertex. Separate photons are labeled with the corresponding numbers in the particle flow diagram. The color coding in both (a) and (b) are to separate different reconstructed particles similar as used in panel (c) of figure~\ref{fig:illustration_pattern_recognition}. The reconstructed particle flow information is also displayed (Refer to the caption of figure~\ref{fig:illustration_pattern_recognition} for more details).}
  \label{fig:pio_reco}
\end{figure}

For a $\pi^0$ produced in a neutral current interaction (or outside the TPC active volume),  there may not be any other TPC information to identify the $\pi^0$ vertex. In these cases, the reconstructed primary neutrino vertex is erroneously connected to an EM shower. A dedicated algorithm is therefore implemented to find a $\pi^0$ with a displaced vertex. For each EM shower, a PCA is performed to find the EM shower's primary axis (i.e. its direction).
Given any two EM showers with calculated primary axes, we find two points, one on each of them. These points are the closest to the opposite primary axis. The midpoint of the pair is then identified and labeled as the displaced vertex.
Once the displaced vertex is determined, the direction of each EM shower is redefined with respect to that vertex for the invariant mass calculation. Figure~\ref{fig:pio_reco}a shows such an example. 

\begin{figure}[H]
  \centering
\begin{overpic}[width=0.60\textwidth]{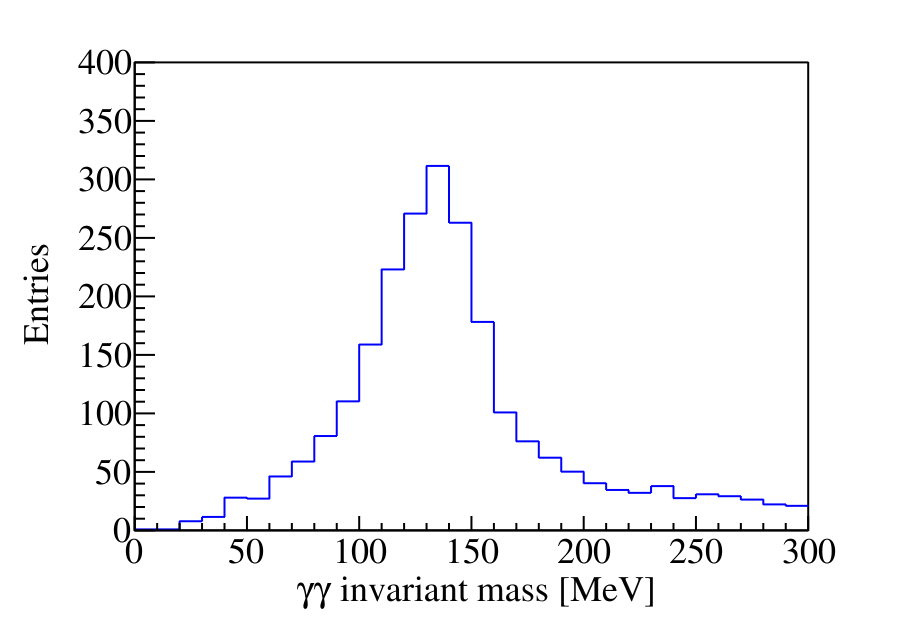}
   \put(52,64){\text{\small MicroBooNE simulation}}
\end{overpic}    
  \caption{
  Reconstructed $\gamma\gamma$ invariant mass ($\sqrt{2E_{\gamma1}E_{\gamma2}\cdot(1-{\rm cos}\theta_{\gamma\gamma})}$) on simulated fully contained charged-current $\nu_\mu\pi^0$ events. The tail at the high reconstructed mass is the result of incorrect association of $\gamma$ candidates. 
  }
  \label{fig:pio_mass}
\end{figure}

For each vertex with two or more associated EM showers, a $\pi^0$ is reconstructed with two separate strategies. The first is to find the pair of photons (above a certain energy threshold) that reconstruct as a  $\pi^0$ with an invariant mass closest to the $\pi^0$ mass. The second is to find the pair of photons that have the highest total energy. 
Both strategies are implemented and applied to all events with two or more photons reconstructed from the same vertex.
The first strategy is used to construct the particle 
flow and is generally good for rejecting backgrounds of $\nu_e$CC events, as incorrectly-paired photons would generally result in an incorrect invariant $\pi^0$ mass. 
The second strategy is used to select $\pi^0$ events, which is useful in calibrating the 
electromagnetic energy reconstruction, given that it has no bias in the invariant mass reconstruction. Figure~\ref{fig:pio_reco} shows two examples of $\pi^0$ reconstructions. In figure~\ref{fig:pio_reco}b the $\pi^0$ vertex is also associated with other activities. Figure~\ref{fig:pio_mass} shows the reconstructed $\pi^0$ mass in simulation for the fully contained $\nu_\mu$CC $\pi^0$ selection. Here, the fully contained events are defined to be events with the reconstructed TPC activity fully contained within the fiducial volume (3 cm inside the effective TPC boundary~\cite{Abratenko:2021bzb}, which is the corrected boundary that takes a space charge effect~\cite{Adams:2019qrr,Mooney:2015kke} into account). Compared to the earlier MicroBooNE results~\cite{MicroBooNE:2019rgx}, the resolution of the reconstructed $\pi^0$ mass is improved by about 15\%.
  
\subsection{Reconstruction Performance of Primary Leptons}~\label{sec:perf_lepton}
The reconstruction quality of the primary leptons (muons and electron EM showers) is essential for both neutrino event flavor tagging and neutrino energy reconstruction. Therefore, we evaluate the reconstruction efficiency and angular resolution. The first step of the evaluation is to define truth-reco matching. We select events with a good reconstructed neutrino vertex (within 1 cm of the truth vertex), to decouple the evaluation of single particle reconstruction and neutrino vertex reconstruction. Then within the selected events, using the simulation truth information, we find all the leading (with largest energy) primary (from neutrino interaction vertex) muons or electrons originating inside a fiducial volume (20 cm from the active volume boundary) as the \textit{truth target}. The fiducial volume requirement ensures that the particles deposit enough visible information for reconstruction in the detector. With the particle flow information reconstructed from the aforementioned pattern recognition algorithms, we find all the leading primary muons or electrons. If those muons or electrons have a reconstructed vertex within 1~cm of the truth counterpart, we count this as a truth-reco matching in the following evaluations in this subsection.

\begin{figure}[!hbt]
		\begin{center}
\begin{overpic}[width=0.49\columnwidth]{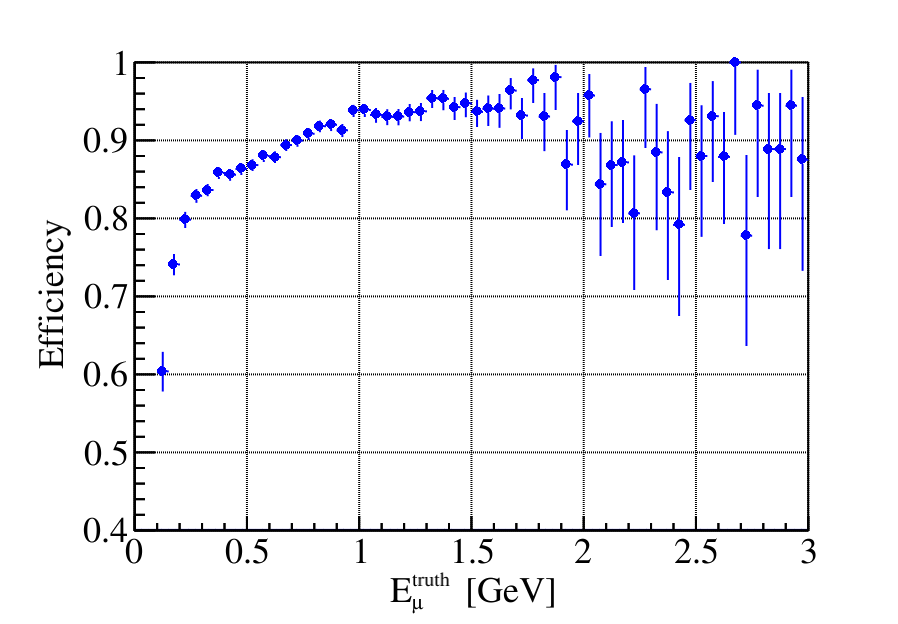}
   \put(42,64){\text{\small MicroBooNE simulation}}
\end{overpic}  
\begin{overpic}[width=0.49\columnwidth]{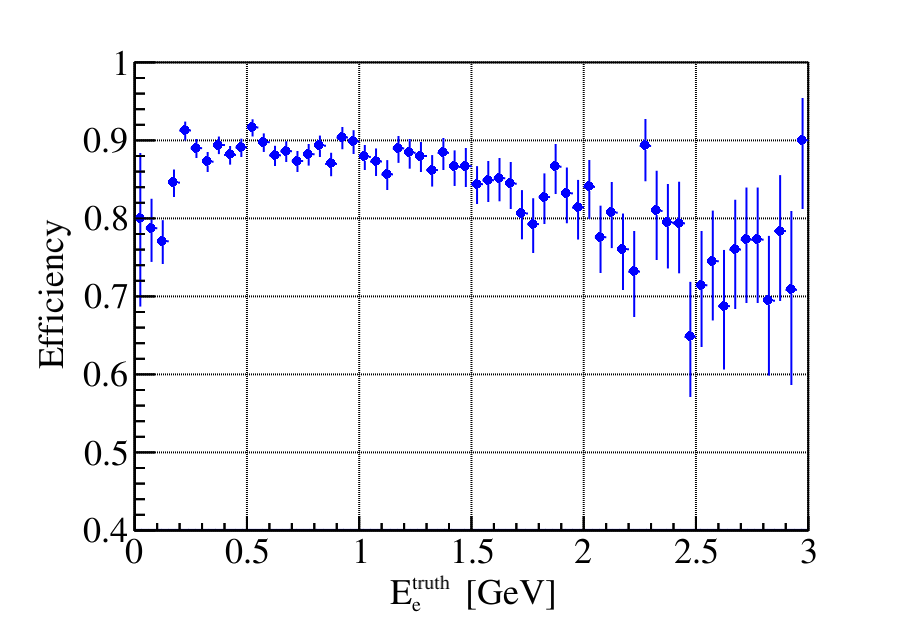}
   \put(42,64){\text{\small MicroBooNE simulation}}
\end{overpic}  
		\caption{Single particle reconstruction efficiency for the leading primary muon (left) and electron (right). The x-axis is the truth particle energy.}
		\label{fig:effSingleParticle}
		\end{center}
\end{figure}

Figure~\ref{fig:effSingleParticle} shows the reconstruction efficiencies as a function of the truth particle energy for leading primary muons and electrons.
These efficiencies are related to the performances of algorithms introduced in this section as well as algorithms from section~\ref{sec:pr_to_vertex}, i.e., the track segment finding, the trajectory and $dQ/dx$ fitting, the track-shower separation, the PID and the neutrino vertex identification.
The denominators are the number of the truth targets aforementioned and the numerators are the number of matched reconstructed counterpart. The muon reconstruction efficiency rises to its plateau at about 85\% to 95\% above 300 MeV. The electron reconstruction efficiency plateaus at about 90\% and slightly decreases above 1.5 GeV. The slight reduction in efficiency at high energy is the result of incorrect reconstruction for isochronous event topologies (close to parallel to the wire planes), where a large amount of ionization electrons arrive at the anode plane at the same time creating ambiguities in the projective wire-plane readout.

\begin{figure}[!hbt]
		\begin{center}
\begin{overpic}[width=0.49\columnwidth]{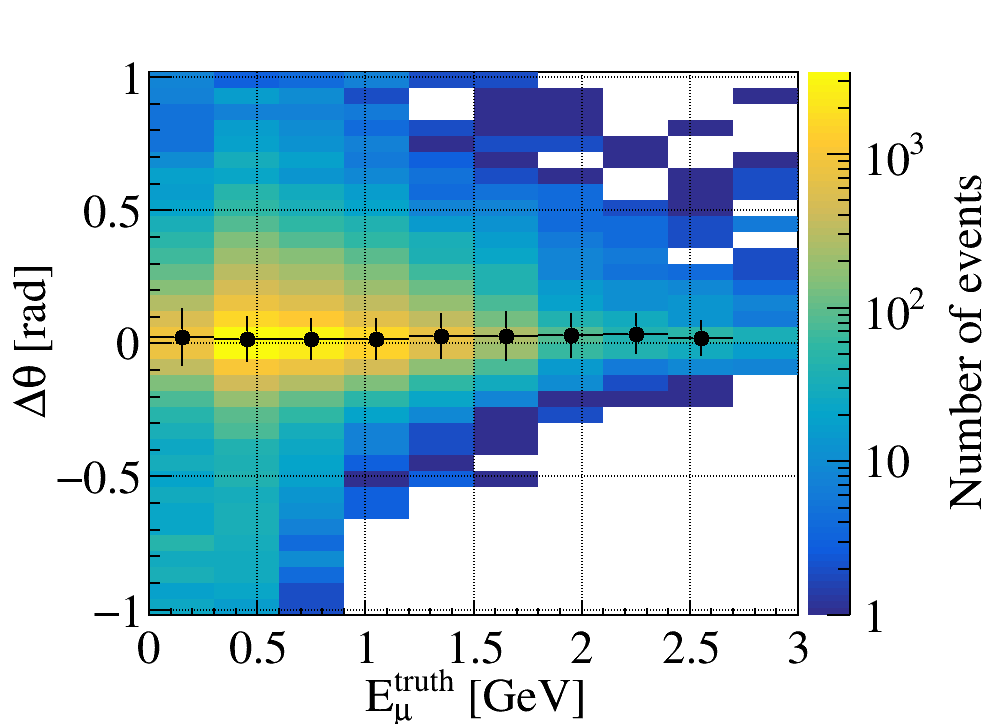}
   \put(32,67){\text{\small MicroBooNE simulation}}
\end{overpic}  
\begin{overpic}[width=0.49\columnwidth]{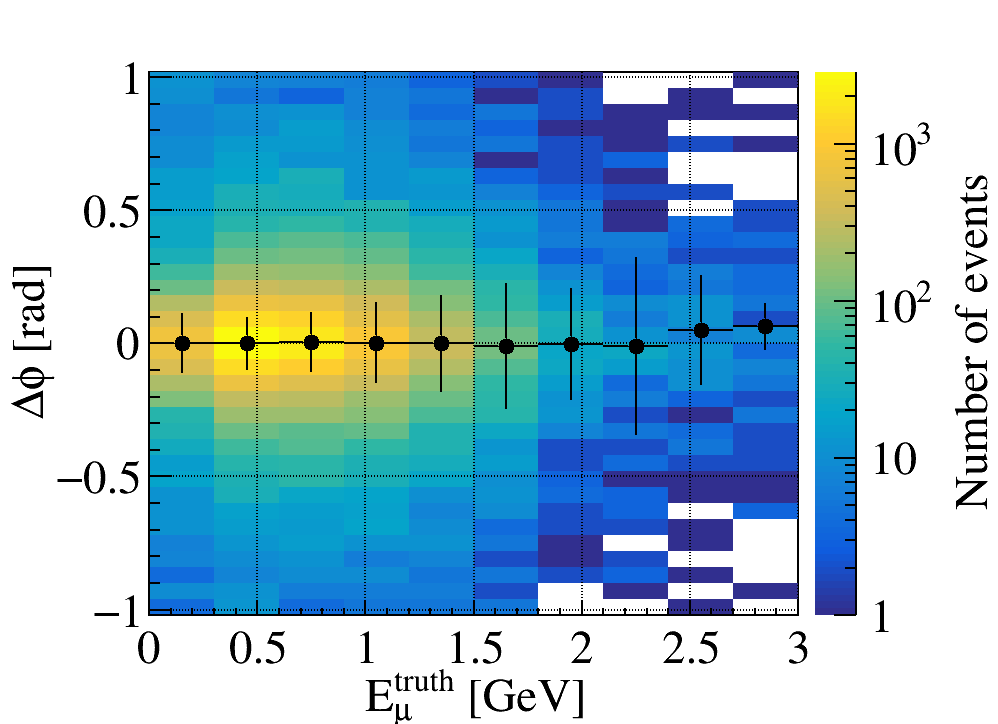}
   \put(32,67){\text{\small MicroBooNE simulation}}
\end{overpic} 
\begin{overpic}[width=0.49\columnwidth]{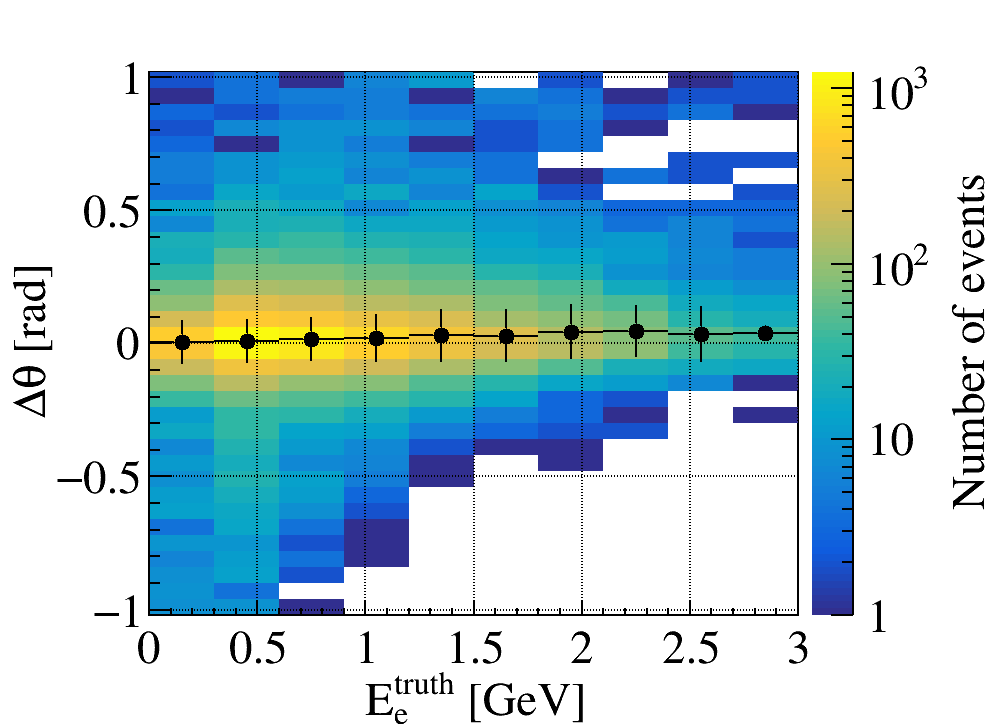}
   \put(32,67){\text{\small MicroBooNE simulation}}
\end{overpic}  
\begin{overpic}[width=0.49\columnwidth]{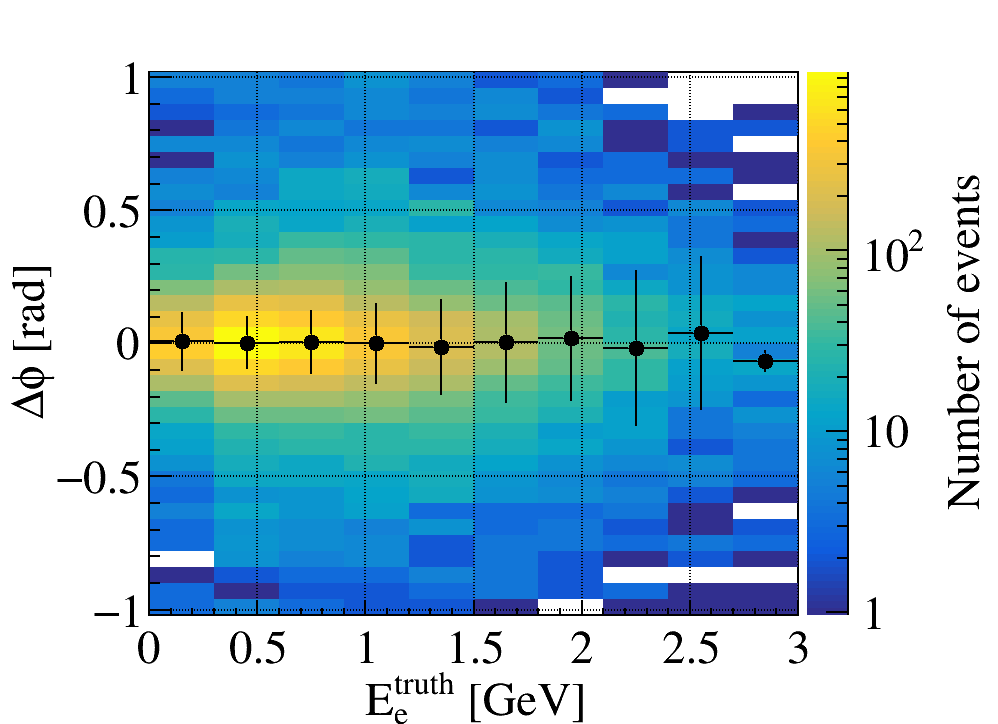}
   \put(32,67){\text{\small MicroBooNE simulation}}
\end{overpic}
		\caption{Difference between reconstructed and truth track angle as a function of truth particle energy for leading primary muons (top) and electrons (bottom). The left panels are for polar angles and the right panels are for azimuthal angles. Slice-wise mean and resolution values from Gaussian fittings are overlaid. The slight bias in the polar angle is because of its non-negative nature in its reconstruction. }
		\label{fig:angleSingleParticle}
		\end{center}
\end{figure}

Figure~\ref{fig:angleSingleParticle} shows the resolution of the reconstructed angle as a function of the truth particle energy 
for leading primary muons and electrons.
Here the $\theta$ and $\phi$ refer to polar and azimuthal angles defined with 
respect to the neutrino beam direction. The truth-reco matching is defined as before.
The angular resolutions for both muon and electron reconstruction are at the 0.1 rad (5.7 degrees) level.

\section{Neutrino Energy Reconstruction}~\label{sec:energy_reco}
With a particle flow tree reconstructed from the neutrino vertex, the neutrino energy can be calculated. There are, in general, three methods to compute the deposited energy of a charged track or shower:
\begin{itemize}
    \item Range: the travel range is used to calculate the energy of track-like particles if they stop inside the detector. The NIST PSTAR database~\cite{pstar} is used to derive the relation between the range and the kinetic energy of different particles with different masses. 
    \item Summation of $dE/dx$: the (visible) kinetic energy is estimated by summing up $dE/dx$ for each piece ($\sim$6~mm) of the track. The energy loss per unit length $dE/dx$ is converted from the ionization charge per unit length $dQ/dx$ considering the recombination effect when the ionization electrons are produced.
    Note that an {\it effective} recombination model that takes into account the overall normalization difference between MicroBooNE data and simulation charge signals~\cite{Adams:2019ssg} is built by tuning the parameters of the modified box model from ArgoNeuT~\cite{ArgoNeuT:2013kpa}. This effective recombination model improves the agreement with MicroBooNE data; however, it still underestimates the energy by about 10\%. 
    \item Overall charge-energy scaling: while the previous two methods are appropriate for track-like objects, they are not suitable for EM showers because of the difficulties in deriving $dQ/dx$ or range. After multiplying 23.6~eV per ionization pair~\cite{Shibamura:1975zz,Miyajima:1974zz}, the energy of EM showers is estimated by applying another overall scaling factor 2.50 ($=1/0.4$). This factor is derived from a simulation that includes the bias in the reconstructed charge~\cite{Abratenko:2020hpp} and the average recombination factor ($\approx$0.5) of an EM shower in argon. For $\nu_e$CC events, the energy reconstruction of the primary electromagnetic shower gives a 12\% energy resolution with about 2\% bias, indicating a good performance of charge clustering for EM showers (described in section~\ref{sec:em_clustering}).
\end{itemize}
The three methods are combined in the neutrino energy estimation.
For a stopped track longer than 4 cm, range method is used.
For short stopped tracks ($<$ 4 cm), the summation of $dE/dx$ method is used, given the range measurement is expected to have a larger uncertainty since track trajectories consist of points separated by 0.6 cm.
For a long muon, where delta rays become visible (multiple track segments are part of a reconstructed muon track), we again use the summation of $dE/dx$ method, given the track length of delta rays may bias the range calculation of the long muon.
For EM showers, the charge scaling method is used, since track trajectories are not well defined in this case.
The neutrino energy is estimated by summing up the kinetic energy of all particles in the reconstructed particle flow tree. For each muon or charged pion or electron, its mass is added to the energy reconstruction. In addition, an average binding energy of 8.6 MeV per nucleon in which case an argon-40 nucleus is completely disassembled is added for each identified proton in the reconstructed particle flow. These protons may be produced at the primary neutrino interaction or secondary interactions (e.g. produced by a neutron).



\begin{figure}[htp]
  \centering
  \begin{overpic}[width=0.49\figwidth]{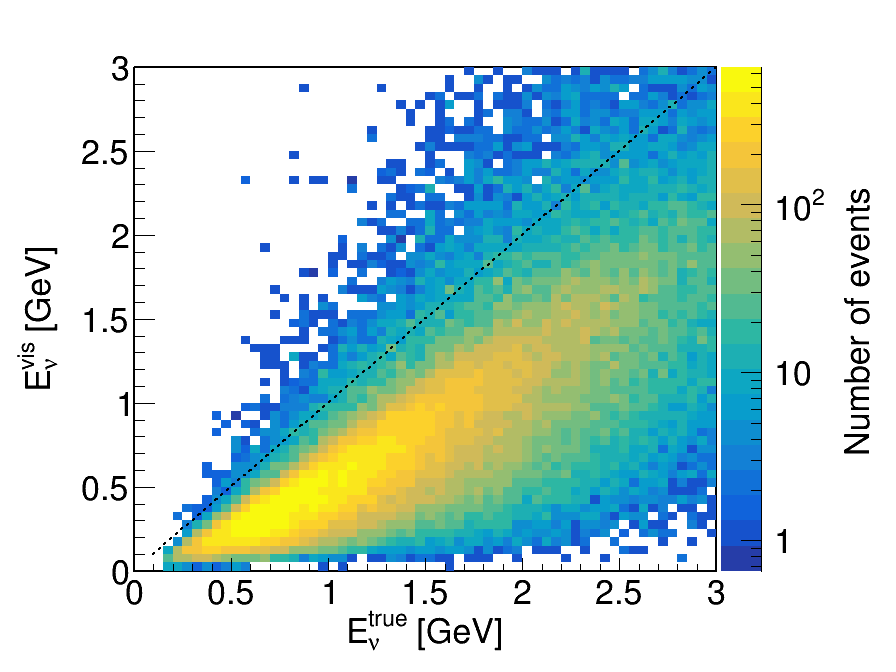}
        \put(29,70){\text{\small MicroBooNE simulation}}
  \end{overpic}  
  \begin{overpic}[width=0.49\figwidth]{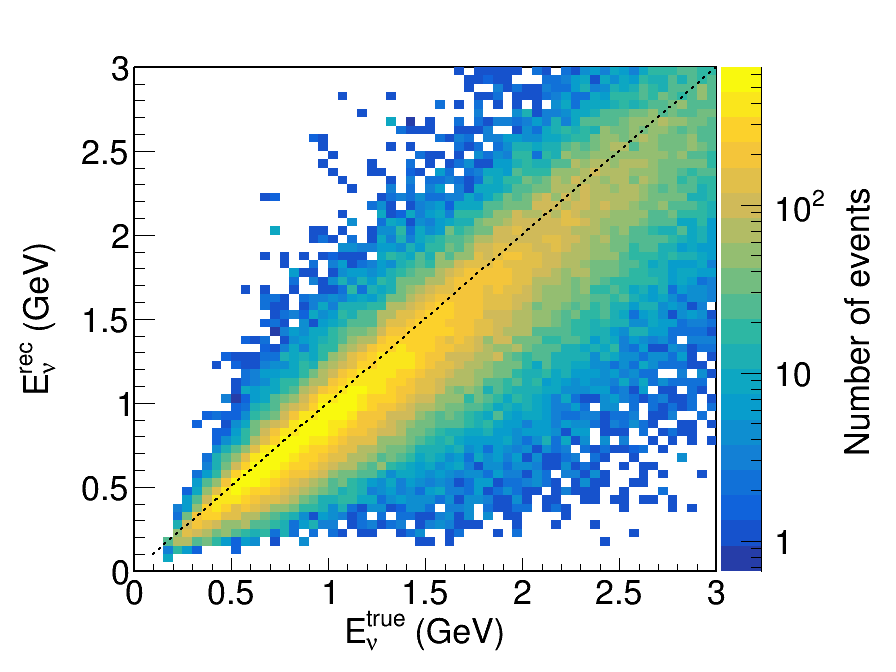}
        \put(29,70){\text{\small MicroBooNE simulation}}
  \end{overpic}  
  \begin{overpic}[width=0.49\figwidth]{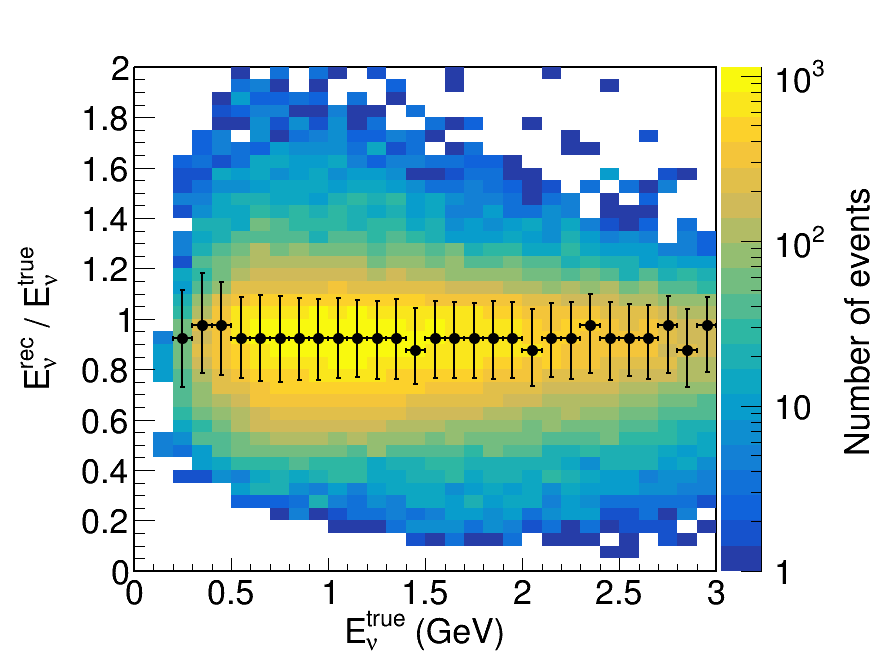}
        \put(29,70){\text{\small MicroBooNE simulation}}
  \end{overpic}    
  \caption{
  Energy reconstruction for selected fully contained $\nu_e$CC candidates: reconstructed visible energy (top left), reconstructed neutrino energy (top right), and ratio of reconstructed neutrino energy to truth (bottom). In the bottom panel, the peak values and the corresponding resolutions (asymmetric) for each true energy bin are plotted as well.
  }
  \label{fig:energy_nueCC_3panel}
\end{figure}

\begin{figure}[htp]
  \centering
  \begin{overpic}[width=0.49\figwidth]{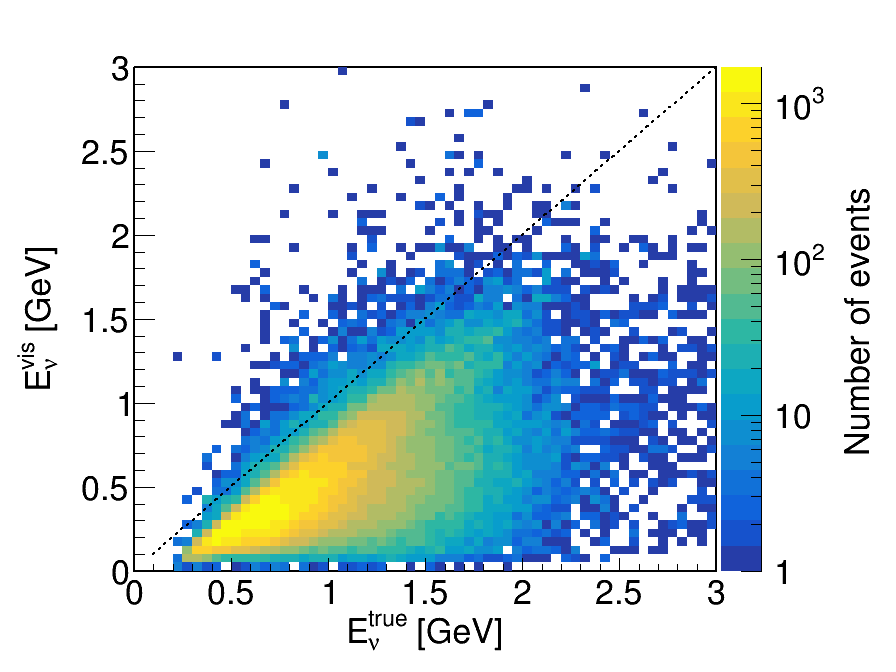}
        \put(29,70){\text{\small MicroBooNE simulation}}
  \end{overpic}  
  \begin{overpic}[width=0.49\figwidth]{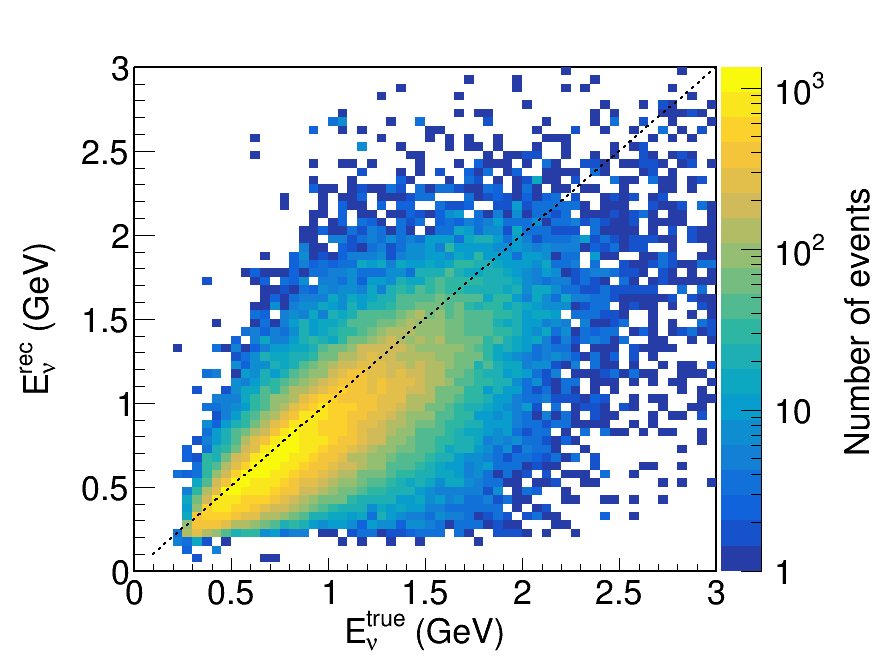}
        \put(29,70){\text{\small MicroBooNE simulation}}
  \end{overpic}  
  \begin{overpic}[width=0.49\figwidth]{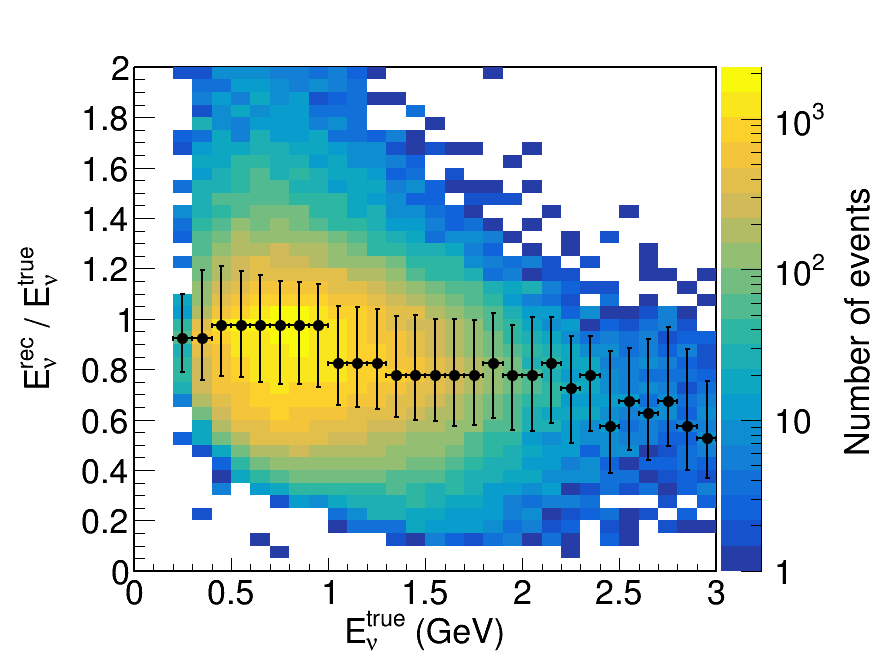}
        \put(29,70){\text{\small MicroBooNE simulation}}
  \end{overpic}    
  \caption{
  Energy reconstruction for selected fully contained $\nu_\mu$CC candidates: reconstructed visible energy (top left), reconstructed neutrino energy (top right), and ratio of reconstructed neutrino energy to truth (bottom). In the bottom panel, the peak values and the corresponding resolutions (asymmetric) for each true energy bin are plotted as well.  
  }
  \label{fig:energy_numuCC_3panel}
\end{figure}

Figures~\ref{fig:energy_nueCC_3panel} and \ref{fig:energy_numuCC_3panel} summarize the neutrino energy reconstruction performance for fully contained $\nu_e$CC and $\nu_\mu$CC interactions.
The top left panels show the reconstructed visible energy and the top right panels show the reconstructed neutrino energy.
We can see that the reconstructed neutrino energy is closer to the diagonal line, meaning that the bias is largely corrected from the reconstructed visible energy.
The bottom panels show that the reconstructed neutrino energy resolution is 10\% to 15\% for $\nu_e$CC events and 15-20\% for $\nu_\mu$CC events.
At a truth neutrino energy of 800 MeV, the reconstructed neutrino energy resolution is 15\% for $\nu_e$CC and 20\% for $\nu_\mu$CC. The bias of the reconstructed neutrino energy relative to the true neutrino energy is 7\% and 10\% on average for BNB $\nu_e$CC and $\nu_\mu$CC interactions, respectively. The bias in the high energy region, e.g. true neutrino energy greater than 1 GeV, is obvious for the selected $\nu_\mu$CC sample. This is because the high energy $\nu_\mu$CC interactions after the selection, in particular with the full containment requirement, are more likely 1) to have high energy transfer (difference of incoming neutrino energy and outgoing lepton energy) which results in a sizable amount of invisible energy carried away by final state hadrons especially neutrons or other neutral particles; and 2) to have high energy muons (long muons with significant delta-rays), as discussed above, the reconstructed energies of which have a relatively large bias because of the usage of an imperfect calorimetric reconstruction method instead of the range method.

In real data analysis~\cite{WC_eLEE, WC_numuXS}, the various methods used to reconstruct neutrino energy are validated. 
The summation of $dE/dx$ method is validated by comparing the $dQ/dx$ vs. residual range for the stopped muons and protons between data and MC.
The energy reconstruction of EM showers is validated by comparing the reconstructed $\pi^0$ mass between data and MC.
The reconstructed hadronic energy has also been validated by comparing the lepton-kinematics-constrained MC prediction and real data measurement, in which case the systematic effect, in particular from the neutrino-argon interaction modeling, is largely reduced.
 
\section{Summary}~\label{sec:summary}
To summarize, we describe a set of pattern recognition algorithms to reconstruct neutrino events based on the Wire-Cell
reconstruction paradigm. These algorithms perform well on the tasks of neutrino vertex identification, single particle
reconstruction, and neutrino energy reconstruction, which are essential to achieve high-performance inclusive $\nu_e$CC
and $\nu_\mu$CC event selections~\cite{WC_eLEE, WC_numuXS}.
Evaluated using the current state-of-the-art MicroBooNE MC simulation, we achieve 65.8\% and 72.9\% neutrino vertex 
identification efficiencies for $\nu_e$ and $\nu_\mu$ charged-current interactions, respectively.
These neutrino vertex identification efficiencies are comparable to those achieved with the Pandora reconstruction paradigm~\cite{Acciarri:2017hat}.
In addition, 80-90\% single particle reconstruction efficiencies for primary leptons and 15-20\% reconstructed 
neutrino energy resolutions are achieved for these charged-current neutrino interactions.
The achievements reported in this paper build a solid foundation for downstream physics analyses, such as the search for new physics with electron neutrinos~\cite{WC_eLEE} and measurement of charged-current muon neutrino interaction cross section on Argon~\cite{WC_numuXS}.
Regarding the performance comparison between the Wire-Cell 
reconstruction and other reconstruction paradigms in MicroBooNE, more information can be found in 
Ref.~\cite{MicroBooNE:2021rmx}, which summarizes the searches for an excess of electron neutrino interactions 
in MicroBooNE using multiple final state topologies and multiple event reconstruction paradigms. 
In addition to the search in the more demanding inclusive $\nu_e$CC channel using Wire-Cell~\cite{WC_eLEE}, 
searches in exclusive quasi-elastic $\nu_e$CC channel~\cite{MicroBooNE:2021jwr} and semi-inclusive $\nu_e$CC 
channel without pions in the final state~\cite{PeLEE_PRD} are performed with the Deep-Learning based 
reconstruction~\cite{MicroBooNE:2018kka,MicroBooNE:2020yze,MicroBooNE:2020hho} and the Pandora
reconstruction~\cite{Acciarri:2017hat}, respectively. 
The current Wire-Cell pattern recognition relies on many traditional (or non-Deep-Learning) algorithms, 
in order to mitigate the risk of amplifying any discrepancies between simulation and real data. As the 
simulation for LArTPCs becomes more mature, we are expecting to adopt more Deep-Learning based algorithms 
in the Wire-Cell reconstruction paradigm to further enhance the overall performance. 


\acknowledgments
This document was prepared by the MicroBooNE collaboration using the resources of the Fermi National Accelerator Laboratory (Fermilab), a U.S. Department of Energy, Office of Science, HEP User Facility. Fermilab is managed by Fermi Research Alliance, LLC (FRA), acting under Contract No. DE-AC02-07CH11359.  MicroBooNE is supported by the following: the U.S. Department of Energy, Office of Science, Offices of High Energy Physics and Nuclear Physics; the U.S. National Science Foundation; the Swiss National Science Foundation; the Science and Technology Facilities Council (STFC), part of the United Kingdom Research and Innovation; the Royal Society (United Kingdom); and The European Union’s Horizon 2020 Marie Sklodowska-Curie Actions. Additional support for the laser calibration system and cosmic ray tagger was provided by the Albert Einstein Center for Fundamental Physics, Bern, Switzerland. We also acknowledge the contributions of technical and scientific staff to the design, construction, and operation of the MicroBooNE detector as well as the contributions of past collaborators to the development of MicroBooNE analyses, without whom this work would not have been possible.

\clearpage  
\appendix

\clearpage
\bibliographystyle{hunsrt}
\bibliography{wc-pat-rec}{}

\end{document}